\documentclass[preprint]{./aastex}
\usepackage{graphicx}
\usepackage{lscape}
\usepackage{color}
\shorttitle{HD 134439 and 134440}
\shortauthors{Chen et al.}

\begin{document}

\title{The Origin of the Metal-Poor Common Proper Motion Pair HD 134439/134440:  Insights from New Elemental Abundances}

\author{Yu Chen and Jeremy R. King}
\affil{Department of Physics and Astronomy, 118 Kinard Lab, Clemson University, Clemson, SC 29634-0978}
\email{yuc@clemson.edu {\ }jking2@clemson.edu}

\and
\author{Ann M. Boesgaard\altaffilmark{{\dagger}}}
\affil{Institute for Astronomy, 2680 Woodlawn Drive, Honolulu, HI 96822}
\email{boes@ifa.hawaii.edu}
\altaffiltext{{$\dagger$}}{Visiting Astronomer, W.M KECK Observatory, which is operated as a scientific 
partnership among the California Institute of Technology, the University of California and the National 
Aeronautics and Space Administration.  The Observatory was made possible by the generous financial support 
of the W.M. Keck Foundation.} 

\begin{abstract}
The low [$\alpha$/Fe] ratio in the metal-poor ([Fe/H]$\sim -1.50$) common proper motion pair HD 134439 and
HD 134440 has been variously attributed to chemical evolution in an extragalactic environment with an
irregular star formation history, planetessimal accretion, and formation in an environment with an
unusually high dust-to-gas ratio.   We explore these various putative origins using CNO, Be, Ag, and Eu 
abundances derived from high-resolution near-UV Keck/HIRES spectroscopy.   While we confirm a 
previously suggested correlation between elemental abundance ratios and condensation temperature 
at the 95\% confidence level, these ratios lie within the continuum of values manifested by extant dSph data.   
We argue that the most plausible origin of our stars' distinctive abundance distribution relative to the Galactic halo 
field is formation in an environment chemically dominated by products of Type II SN of low progenitor mass; 
such a progenitor mass bias has been previously suggested as an explanation of low $\alpha$-element 
ratios of dSph stars.  The proper motion pair's heavy-to-light 
$n$-capture element ratio, which is ${\ge}0.3-0.5$ dex lower than in the Galactic halo field and dSph stars, 
is discussed in the context of the truncated $r$-process, phenomenlogical $n$-capture production models, and $\alpha$-rich freezeout in a high neutron excess environment;  the latter simultaneously provides an attractive explanation of the difference in 
[Ca,Ti/O,Mg,Si] ratio in HD 134439/134440 compared to ${\it in{\ }situ}$ dSph stars.   
\end{abstract}


\section{INTRODUCTION}
 
The outcome of the \citet{1962ApJ...136..748E} rapid monolithic collapse (ELS) model 
is an old metal-poor Galactic halo with enhanced [$\alpha$/Fe] values, and a young metal-rich 
disk with near-solar [$\alpha$/Fe].  In this picture, stellar [$\alpha$/Fe] values diminish 
as a function of Galactic time as metallicity increases from nucleosynthetic feedback 
that is a sensitive function of stellar mass.  However, stars in the Galaxy do not strictly 
follow the simple trend expected by ELS model.  \citet{1997AJ....114..363C} reported the low 
metallicity ([Fe/H]$=-1.86$) star BD+80$^{\circ}$245 to have an abnormally low average 
[$\alpha$/Fe]${\sim}-0.3$.  CS 22966-043, a blue metal-poor SX Phe variable star, exhibits 
unusual $\alpha$-element deficiencies \citep{2000AJ....120.1014P}.  Other examples include 
\citeauthor{1997A&A...326..751N}'s \citeyearpar{1997A&A...326..751N} sample of low-$\alpha$ 
halo stars with large apogalactocentric distances, the expanded \citet{NS10} sample of 
low-$\alpha$ halo stars that are weakly bound and evince predominantly retrograde orbits, the 
young globular clusters Rup 106 and 
Pal 12 \citep{1997AJ....114..180B}, and the halo field common proper motion pair HD 134439 and 
134440 \citep{1997AJ....113.2302K}.  

The early ELS picture of Galactic formation is undoubtedly much too simplistic, lacking consideration 
of a variety of important feedback mechanisms associated with star formation \citep[e.g.,][]{Hopkins13a}, 
the effects of recycling timescale \citep[e.g.][]{DeLucia14}, the context of  cosmological influences of dark matter and energy on Galaxy formation \citep[e.g.,][]{Penzo14}, 
and the consequences of rapid infall of cool intergalactic gas \citep[e.g.,][]{Hopkins13b}.  Regardless, the existence of low-$\alpha$ stars suggests that an additional or alternative formation mechanism of 
the Galactic halo is required to explain these abundance patterns. 

By examining color-magnitude diagrams of globular clusters in the Galactic halo, 
\citet{1978ApJ...225..357S} found a diverse horizontal-branch morphology corresponding to 
an age spread of ${\ge}10^{9}$ years, suggesting the halo formation timescale lasted longer 
than a free-fall time.  Consequently, they suggested the outer halo of the Galaxy may be 
a product of accreted fragments that formed independently of the Galactic 
collapse.  Indeed, a variety of surveys evince the presence of significant 
tidal streams \citep{2001ApJ...551..294I} and substructure \citep{2000ApJ...540..825Y} in the 
halo that is interpreted as detritus from dwarf galaxy encounters 
\citep{2001ApJ...547L.133I,2008ApJ...689..936J}.  Current hydrodynamic simulations of the formation of 
late-type spirals naturally produce stellar halos populated by accreted ex-situ stars 
\citep[e.g.][]{Pillepich14}.  

This hierarchical picture of Galactic development is supported by additional independent evidence.  For 
example, presolar SiC grains show unexpected Si isotopic ratios \citep[see][]{1998AREPS..26..147Z} that  
can be explained by a solar birth environment located in the remains of a merger event with a metal-poor satellite 
galaxy 5-6 Gyr ago \citep{2003ApJ...598..313C}.   Large pre-solar $^{18}$O abundances have also been explained 
as the result of feedback of SNe outbreaks caused by the merger of a satellite galaxy with the Milky 
Way \citep{2004LPI....35.1045C}.  \citet{2007EAS....24....3P} shows that the halo metallicity 
distribution is well reproduced by the merger of a collection of sub-halo satellite systems 
having various masses.  

The properties of present-day dwarf spheroidal (dSph) systems provide evidence of hierarchical 
formation of the Galactic halo \citep{2002ApJ...575...18M}.  dSphs have distinctly 
different environments than the Milky Way--notably their irregular and/or small star formation 
rates and strong galactic winds that exhaust their gas content quickly \citep{2003MNRAS.345...71L}.  
The low star formation rate, and thus slow metal enrichment, in dSph galaxies results in 
their stellar members possessing elemental abundance ratios similar to more metal-rich Galactic 
stars.  Hence, the metal-poor [$\alpha$/Fe] values in dSphs would be lower than a Galactic halo star of 
the same metallicity.  The contribution of {\it s}-process nuclei, produced primarily in 
long-lived low-to-intermediate mass stars, does not become important until 
[Fe/H]$\sim-1.0$ dex in Milky Way \citep{2001ApJ...559..925Q, 2004ApJ...601..864T}.  However, 
the {\it s}-process production may become significant at lower [Fe/H]$\sim-1.7$ dex in dSph galaxies 
due to the slower metallicity increase \citep{2004AJ....128.1177V, 2008A&A...481..635L}.  These
expectations are seemingly confirmed by a growing body of dSph abundance data seen in, e.g., 
Figures 3, 5, 6, 7, 9, and 16 of \citet{CH2009}; Figures 10 and 14 of \citep{Letarte2010}; and in Figures 
11, 13, and 14 of the outstanding review of \citet{Tolstoy2009}.  

\citet{2002AJ....123.1647S} used high-resolution and high S/N spectroscopy to determine abundances 
in some 50 halo field stars with $-3.6{\le}{\rm [Fe/H]}{\le}-0.7$ with peculiar kinematics 
or distinctive orbital properties.  This sample exhibits a real dispersion in [${\alpha}$/Fe], and 
\citet{2002AJ....123.1647S} find that their outer halo stars evince lower [${\alpha}$/Fe] ratios than inner 
halo stars of the same metallicity.  Despite their kinematic-based selection criteria, these authors 
were unable to identify any new low-$\alpha$ stars and concluded that their sample did not 
harbor the chemical hallmarks of an accreted extragalactic population.  

\citet{2002AJ....123..404F} 
examined the abundances and kinematics of ${\sim}70$ halo field stars with $-2{\le}{\rm [Fe/H]}{\le}-1$ 
in the {\it Hipparcos\/} catalog.  A chemo-kinetic relation is manifest in these stars via 
the correlation between a variety of [X/Fe] ratios (including X$=$Mg, Si, and Ca) and 3-d space 
velocity.  In their analysis of 94 metal-poor dwarfs, \citet{NS10} find their low [$\alpha$/Fe] 
stars show distinctive kinematics, which they interpret as a result of accretion from dwarf galaxies.  
However, \citet{2002AJ....123..404F} have called attention to differences in the abundance ratios of 
Galactic low-$\alpha$ stars and those measured {\it in situ} in local dwarf galaxies, 
suggesting the former do not have their origin in the latter. 

An alternative interpretation of low-${\alpha}$ halo stars is provided by \citet{2003ApJ...598L..47S}, who 
posit photospheric pollution due to accretion of iron-rich planetary 
material that increases the surface Fe abundance relative to the $\alpha$-elements.  
\citet{2001AJ....121.3207S} have noted the existence of Galactic disk exoplanet host stars that 
demonstrate increasing abundance [X/H] with increasing condensation temperature $T_{\rm C}$(X), 
and suggest that this correlation is consistent with photospheric accretion of (presumably once 
circumstellar) refractory material that leads to an abundance-$T_{\rm C}$ trend similar to that 
seen in chondritic solar system material (e.g., Figure 1a of Yin 2005).  Photospheric $T_{\rm C}$(X)-dependent
abundance trends are now commonly utilized as signatures of planet formation 
\citep[e.g.,][]{Ramirez2010,Schuler2011a,Schuler2011b,Liu2014}

\citet{2006MNRAS.370.2091C} called attention to correlations between [X/Fe] values 
and $T_{\rm C}$(X) in the low-${\alpha}$ halo stars HD 134439 and 134440.  They propose that 
such correlations originate from formation in a (chemically fractionated, refractory-rich) 
``dusty environment'' in a dSph.  Indeed, the apparent correlation they identify is a qualitative 
mirror image of the (gas-phase) ISM depletion versus $T_{\rm C}$ trend (e.g., Figure 1b of Yin 2005).   
 
Motivated by the various suggested origins of abundance patterns in low-$\alpha$ Galactic halo stars, 
we continue the delineation of abundances in the two metal-poor ([Fe/H]$\sim-1.50$) low-$\alpha$ halo 
field stars HD 134439 and 134440 \citep{1997AJ....113.2302K}.  These two main sequence dwarfs are a 
weakly bound, high proper motion pair with a common origin evidenced by their shared kinematics and 
chemical abundances.   We use newly-derived and existing literature abundances to explore the origin 
of these stars' low [$\alpha$/Fe].    In particular, we present abundances for the light element Be, the volatile 
CNO elements, and the $n$-capture elements Ag and Eu.  We investigate whether there exists a relationship between elemental 
abundances and condensation temperature in these stars and, if so, its interpretation.  Our central 
aim is to determine whether these 2 stars (and perhaps the Galactic population of low-$\alpha$ stars
as a whole) are best understood in the context of star formation environment, circumstellar 
evolution, stellar evolution, extragalactic chemical evolution, or some combination of these.  

\section{DATA AND ANALYSIS}

\subsection{Observations}

High-resolution ($R=49,000$) spectroscopy of HD 134439 and 134440 was obtained on UT 19 June 2006 
using the HIRES spectrograph on the W.M. Keck I 10-m telescope (Table \ref{tab:observation}). The 
useful wavelength coverage of the echelle spectra extends from 3020 to 5800 {\AA}. Standard 
echelle reductions (debiasing, flat-fielding, order tracing and extraction, and wavelength calibration) 
were carried out.  Certain regions of spectra overlapped in two adjacent orders and were coadded to increase 
S/N ratio. The final average S/N values near the CH, NH, and OH features of interest here are listed in 
Table \ref{tab:observation}.  This per-pixel S/N ratio is calculated from Poisson statistics in the pseudo-continuum 
near the features.   

\subsection{Basic Physical Parameters}

The metallicity ([Fe/H]), surface temperature (T$_{\rm eff}$), and microturbulent velocity ($\xi$) 
of HD 134439 and 134440 and their uncertainties are taken from \citet{1997AJ....113.2302K} and are
listed in Table \ref{tab:parameters}.  Surface gravities taken from a 12 Gyr Yonsei-Yale (Y$^{2}$) 
isochrone \citep{2004ApJS..155..667D} with [$\alpha$/Fe]=0, [Fe/H]$=-1.50$, and the 
\citet{1998A&AS..130...65L} color table are $\log g=4.77$ and 4.79 for HD 134439 (T$_{\rm eff}=5000$ K) 
and HD 134440 (T$_{\rm eff}=4785$ K) respectively.  Alternatively, $\log g$ can be calculated 
according to  
\begin{equation}\label{gravity}
\log\frac{g}{g_{\odot}}=\log\frac{\rm M}{{\rm M}_{\odot}}+
4\log\frac{{\rm T}_{\rm eff}}{{\rm T}_{\rm eff,\odot}}
+0.4{\rm V}_{\circ}+0.4BC+2\log\pi+0.12
\end{equation}
where M is the mass, V$_{\circ}$ is the apparent magnitude, and $\pi$ is the parallax in arcsec; these 
values are given in Table 1.  The bolometric correction ($BC$) is given by 
\begin{equation}
BC={\rm M}_{bol,\odot}-{\rm M}_{\rm V}-2.5 \log\frac{\rm L}{{\rm L}_{\odot}}
\end{equation}
We take M$_{bol,\odot}$=4.71 (from the \citealt{1998A&AS..130...65L} color table according to 
\citealt{2001ApJS..136..417Y}), the Hipparcos-based M$_{\rm V}$ values from Table \ref{tab:observation}, 
and $\log\frac{\rm L}{{\rm L}_{\odot}}$ values of $-0.829$ and $-0.931$ (for HD 134439 and 134440, 
respectively) from the Y$^{2}$ isochrone; equation 2 then yields bolometric corrections of $-0.29$ and 
$-0.35$ mag for HD 134439 and 134440, respectively.  Equation \ref{gravity} with $\log g_{\odot}$=4.44 dex
and T$_{\rm eff,\odot}$=5770 K then yields physical gravities of $\log g= 4.61$ and $4.63$ for HD 134439 
and 134440, respectively.  Averaging the isochrone-based and parallax-based physical $\log g$ values, 
we obtain $\log g=4.69$ and $4.71$ for HD 134439 and 134440; the differences between the isochrone and
physical gravities around these means suggest an uncertainty at the $\pm 0.10$ dex level.

\subsection{Abundance Analysis}

The new abundances presented here are determined using Kurucz model atmospheres and an updated version 
of the LTE spectrum synthesis package MOOG package \citep{1973ApJ...184..839S} that includes updated 
b-f opacity data relevant for the near-UV syntheses.  The line list for the 3167.17 {\AA} OH line region is that 
compiled by \citet{2006AJ....131.1057S}.  Other line lists 
used here are compiled from Kurucz\footnote{http://kurucz.cfa.harvard.edu/linelists.html} atomic and molecular 
lines \citep{1995ASPC...78..205K}, the 
Vienna Atomic Line Database \citep{2000BaltA...9..590K}, CH molecular lines from 
\citet{1996A&A...315..204J}, and molecular lines simulated by LIFBASE \citep{lifbase}.  CH, NH 
and OH diatomic dissociation energies adopted in the line lists are 3.47, 3.45, and 4.39 eV, 
respectively.  Oscillator strengths ($gf$-values) were adjusted to produce solar syntheses 
matching the Kurucz solar flux atlas \citep{2005MSAIS...8..189K}. The line lists are provided in
Tables 3-7, which include wavelengths, species identifications, lower excitation potentials, 
and oscillator strengths. 
 
Synthetic solar flux spectra were generated with elemental abundances  
from \citet{1989GeCoA..53..197A} except for C, N, and O.  The solar C and N abundances are 
from \citet{2005ASPC..336...25A}, while the solar O abundance is from \citet{2001ApJ...556L..63A}.  
We note that these adopted abundances not only set the $gf$ values used in deriving the stellar abundances, but
are also subsequently used to solar-normalize those stellar abundances;  our stellar abundances in the
form [X/H] are thus not dependent on the adopted solar CNO values.  
Elemental abundances from previous analyses (see \S 3) were manual inputs in the stellar 
syntheses; abundances of other elements (except Be, C, N, O, and Ag, whose abundances are determined here) were 
scaled to the solar values according to the [Fe/H] value taken here ($-1.47$ dex and $-1.53$ dex for HD 134439 
and 134440, respectively).  A Gaussian smoothing factor, which accounts for instrumental and any modest 
macroturbulent broadening, was obtained from Doppler-corrected FWHM measurements of weak clean observed 
spectral lines at redder wavelengths (4000-5800 {\AA}), and then applied to the synthetic spectra.  

The stellar spectra were first normalized relative to the pseudo-continuum with SPECTRE 
\citep{1987BAAS...19.1129F}.  Small wavelength shifts and flux rescalings were applied to improve the match 
with the synthetic spectra generated by MOOG.  We determined best-fit abundances via $\chi^{2}$ minimization 
of the synthesized and normalized observed spectra over regions centered on features that were identified as 
sufficiently strong and clean to allow good abundance measurements; we discarded those features that were too 
saturated or severely blended.  After initial CNO abundances were measured, they were used as inputs for new 
syntheses in order to enforce molecular equilibrium.  The abundances reported in Tables \ref{tab:carbon}, \ref{tab:nitrogen}, 
and \ref{tab:oxygen} are final measurements after achieving molecular equilibrium. 

\subsection{Carbon Abundances}

\begin{figure}
\plotone{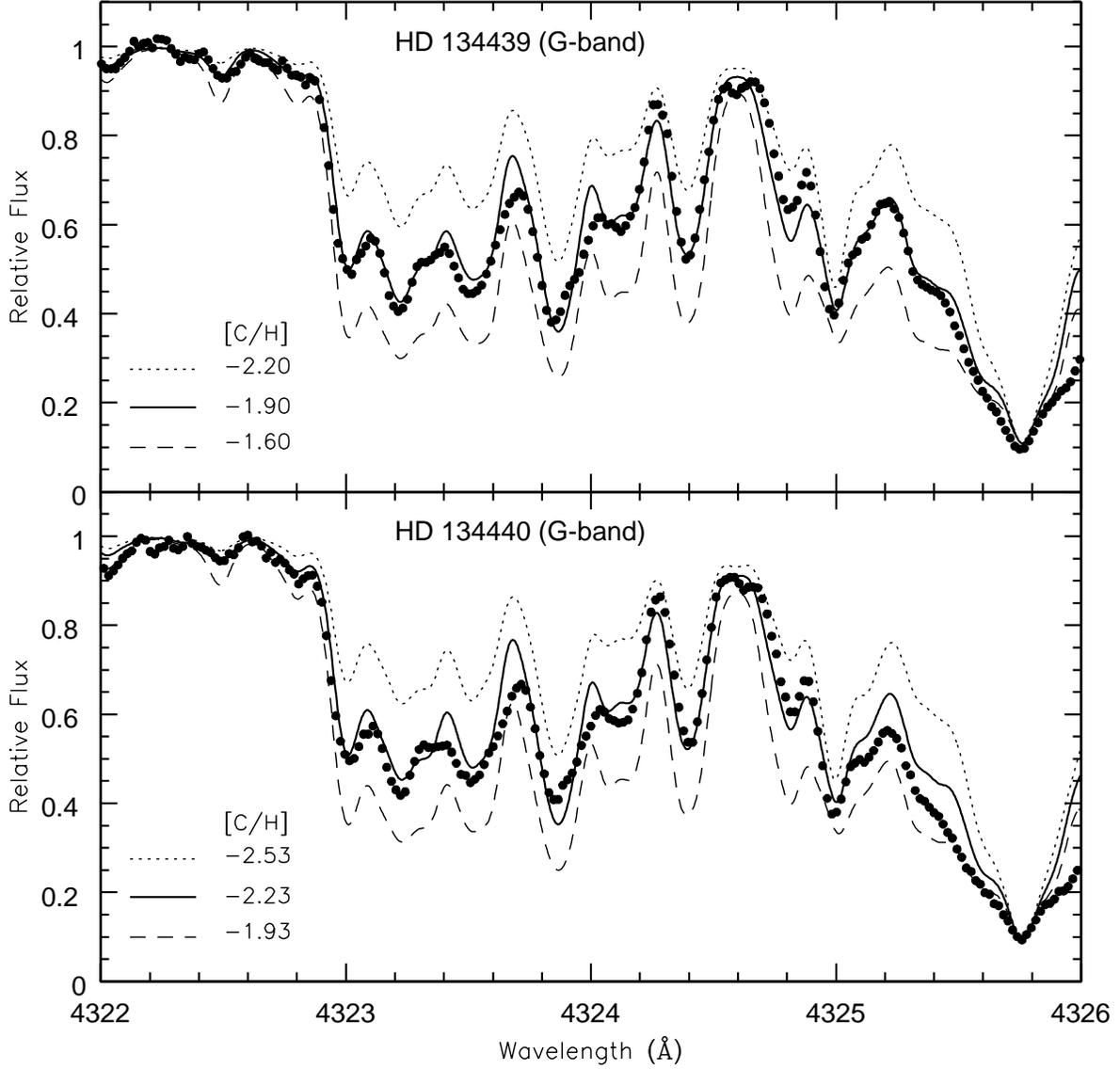}
\caption[G-band (CH) Syntheses]
{G-band synthesis for HD 134439 (upper panel) and HD 134440 (lower panel). 
Filled circles indicate observed spectra. 
The solid line is the average best-fit [C/H] abundance determined by $\chi^{2}$-tests on 
individual lines. 
Dotted and dashed lines are $\pm$0.30 dex deviations from the best-fit abundance.}
\label{fig:carbon}
\end{figure}

Seven CH lines were used in the ${\lambda}$4322-4326 G-band region to 
measure the C abundance (see Figure \ref{fig:carbon}).  The abundances are measured by applying 
$\chi^{2}$ measures to individual features, instead of the entire band.  Final best fit abundances 
are listed in Table \ref{tab:carbon}. The mean carbon abundances found for HD 134439 and 134440 are 
$\log N({\rm C})_{39}=6.49$ and $\log N({\rm C})_{40}=6.16$\footnote{The ``39'' and ``40'' subscripts 
designate HD 134439 and 1344404, respectively; logarithmic number abundances are on the traditional 
scale where log $N$(H)$\equiv$12.}, yielding [C/H]$_{39}=-1.90$ 
for HD 134439 and [C/H]$_{40}=-2.23$ for HD 134440 after self-consistently employing the solar C
abundance used in our line list calibration.

\subsection{Nitrogen Abundances}

\begin{figure}
\plotone{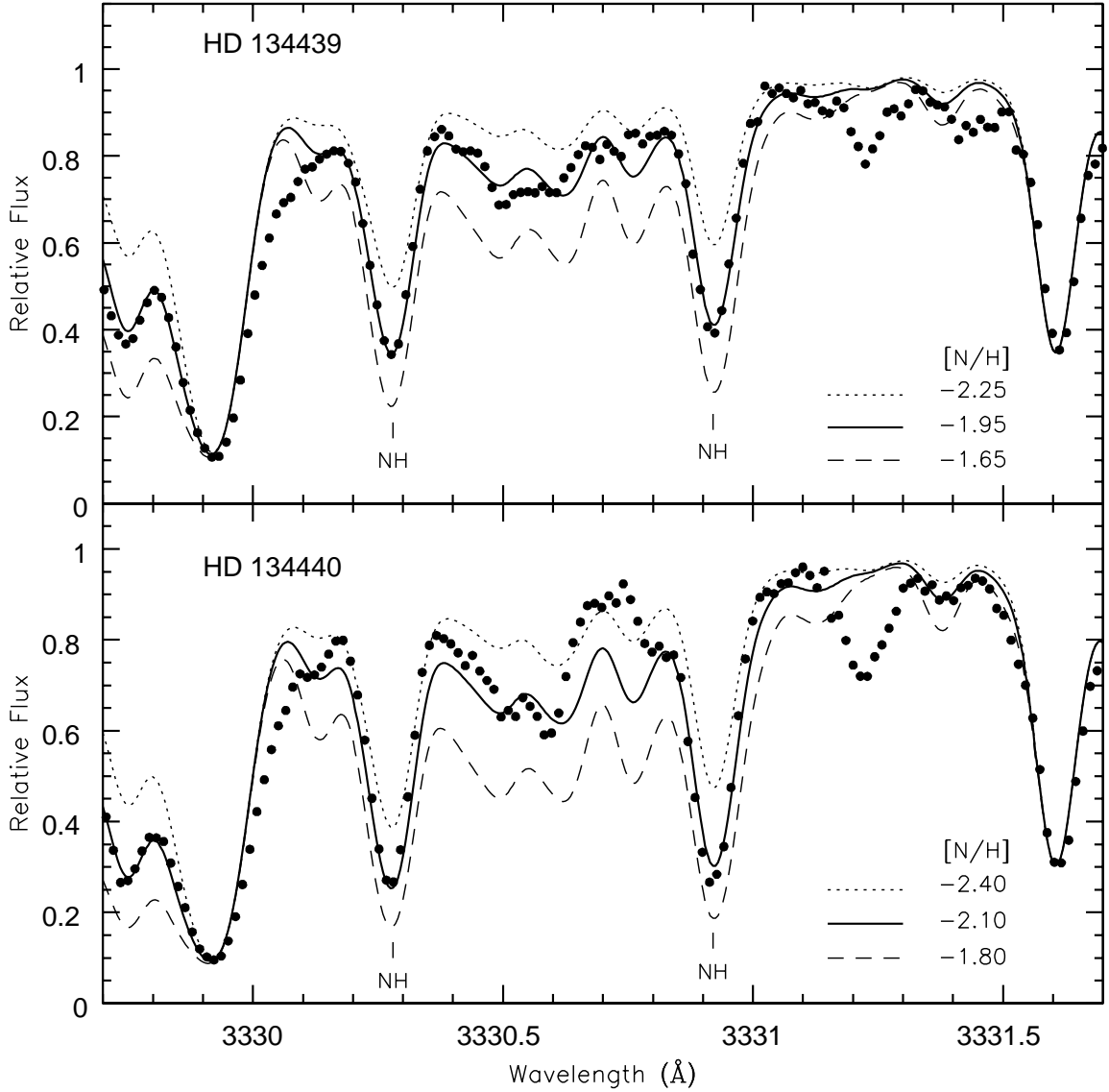}
\caption[NH Syntheses -- 3330{\AA}]
{Synthesis of NH molecular lines for HD 134439 (upper panel) and HD 134440 (lower panel). 
Filled circles indicate observed spectra.
The solid line is the average best-fit [N/H] abundance determined by $\chi^{2}$-test on 
the individual NH lines indicated. 
Dotted and dashed lines are $\pm$0.30 dex deviations from the best-fit abundance.}
\label{fig:nitrogen2}
\end{figure}

Four NH lines in the ${\lambda}$3326-3332 region were used for the calculation of 
the N abundance (see Figure \ref{fig:nitrogen2}).  Although there are more than four NH 
features in the spectral region, we found them too strong or blended relative to the selected 
lines.  The abundances given by the best fit between the synthetic and observed spectra for 
individual NH lines are listed in Table \ref{tab:nitrogen}.  The resulting average nitrogen 
abundances are $\log N({\rm N})_{39}=5.83$ and $\log N({\rm N})_{40}=5.68$.  The ratios relative 
to the Sun are [N/H]$_{39}=-1.95$ and [N/H]$_{40}=-2.10$.

\subsection{Oxygen Abundances}

\begin{figure}
\plotone{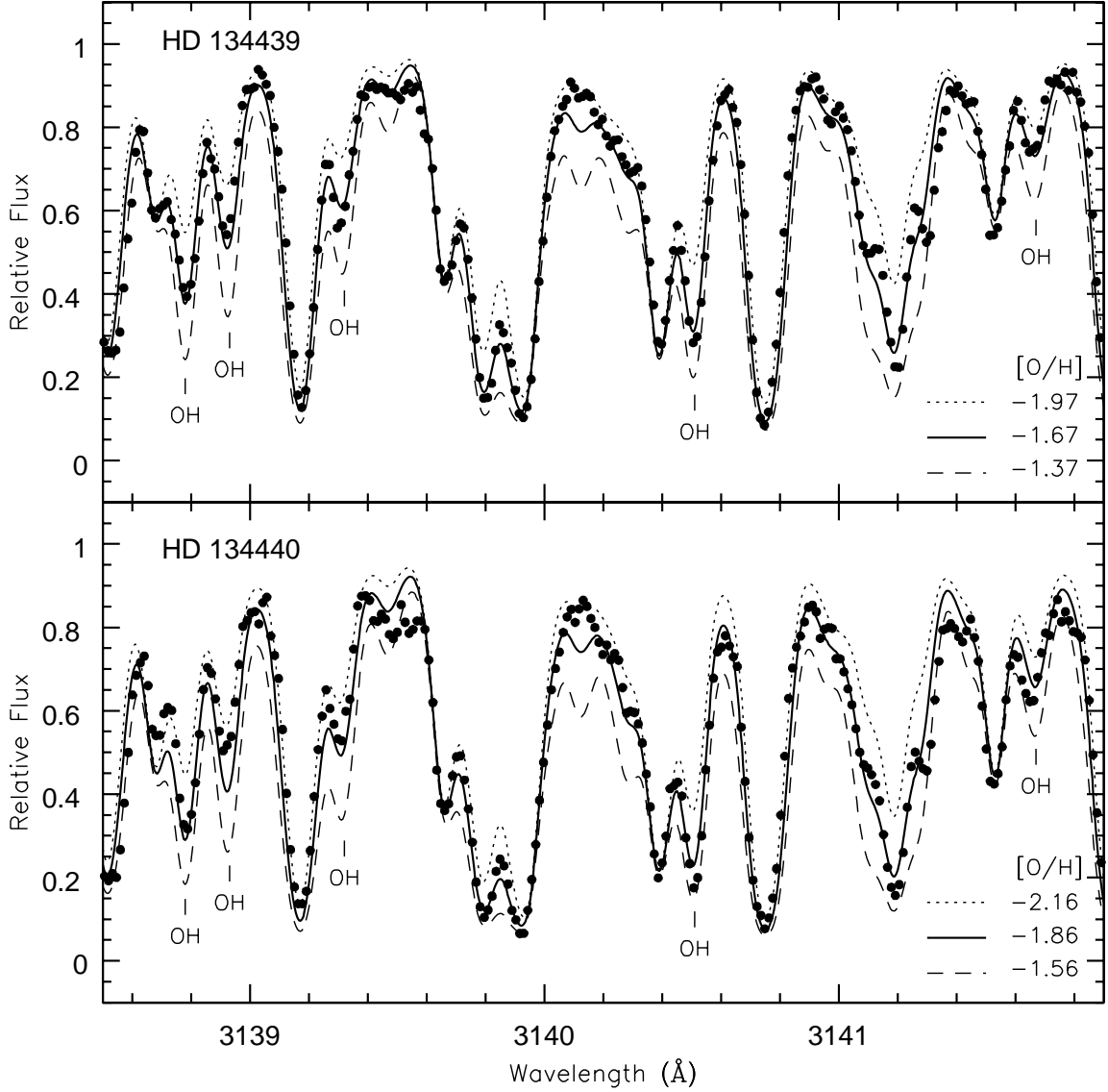}
\caption[OH Syntheses -- 3140{\AA}]
{Synthesis of OH molecular lines for HD 134439 (upper panel) and HD 134440 (lower panel). 
Filled circles indicate observed spectra.
The solid line is the average best-fit [O/H] abundance determined by $\chi^{2}$-test on 
all the OH lines considered. 
Dotted and dashed lines are $\pm$0.30 dex deviations from the best-fit abundance.}
\label{fig:oxygen2}
\end{figure}

The O abundance is measured with seven near-UV electronic OH features in the
range 3129-3168 {\AA} (Figure \ref{fig:oxygen2}).  Line-by-line results are listed in Table \ref{tab:oxygen}. 
Final average oxygen abundances are $\log N({\rm O})_{39}=7.02$ and $\log N({\rm O})_{40}=6.83$. 
The O abundances relative to the Sun are [O/H]$_{39}=-1.67$ and [O/H]$_{40}=-1.86$.

Because the final CNO abundances are normalized with respect to the Sun self-consistently, they 
are independent of adopted $gf$-values and solar abundances.  Utilizing, for 
example, solar abundances different from those employed here would require modifications of the 
$gf$-values in the line lists to match the solar atlas.  Resulting stellar syntheses would produce 
elemental abundances that are adjusted accordingly.  These absolute abundances, however, would then 
be normalized by an equivalently different adopted solar abundance.  Hence, abundances of the elements 
that are consistently normalized to the Sun will be unaffected by the solar values assumed 
in the line list calibration.

\subsection{Abundance Sensitivities and Uncertainties}

Uncertainties in our abundances arise from uncertainties in effective temperature, surface gravity,
microturbulent velocity, the (pseudo-)continuum setting, and measurement uncertainty. 
Uncertainties in the [CNO/Fe] ratios due to uncertainties in the stellar parameters are derived from 
the parameter sensitivities, which are listed in columns 2, 3, and 4 of Table \ref{tab:sensitivities}.  
The sensitivities of [Fe/H], which are folded in to the listed [X/Fe] sensitivities, are taken from Table 3 
of \citet{1997AJ....113.2302K}, who uses the same model atmospheres and is the source of the 
[Fe/H] values and adopted parameters.  

While the observed spectra's (pseudo-)continuum shape and absolute level are difficult to 
determine in isolation, the syntheses provide guidance. Repeating comparison with syntheses 
after altering the continuum level of the observed spectra by plausible amounts, we believe 
the 1$\sigma$ level uncertainties in the average abundance of C, N and O are 0.04 dex 
to 0.06 dex.  The line-by-line spread, parameterized as a statistical uncertainty in the mean 
${\sigma}_{\mu}$, of the C or N or O abundances provides an estimate of 
the measurement uncertainties.  These values for the 7 CH lines, 4 NH lines and 7 OH lines 
are combined in quadrature with the uncertainty in the mean for [Fe/H] \citep{1997AJ....113.2302K},  
yielding the statistical uncertainties in the mean [X/Fe] given in the penultimate column of Table \ref{tab:sensitivities}. 

The total uncertainties  in the [CNO/Fe[ ratios are found by adding the parameter-based, 
continuum-based, and measurement uncertainties in quadrature, and are listed in  listed in the final 
column of Table \ref{tab:sensitivities}.  The major contribution to the final uncertainties originates from 
uncertainties in T$_{\rm eff}$;  our elemental abundances are less sensitive to other quantities, especially $\xi$.  
The mean [CNO/Fe] values in HD 134439 versus HD 134440 differ by 0.09-0.27 dex.  The difference between 
the HD 134439 and 134440 $T_{\rm eff}$ values according to the newer infrared flux 
method results of \citet{Casa2010} is 88 K lower than that associated with our adopted values.  This would diminish
the [CNO/Fe] differences by 0.05-0.09 dex, bringing the [NO/Fe] ratios into near-perfect agreement and leaving a 0.22
dex difference in [C/Fe].   Regardless, our O and N ratios in the two stars differ at the ${\le}1{\sigma}$ level, and 
the 0.27 dex difference in the C ratio is at the $2{\sigma}$ level; a difference of this latter size in one of the 22 elemental 
ratios collected in Table 7 is expected in a statistical sense.  The modest CNO differences between the two stars  do 
not affect the salient result that [CNO/Fe] are all markedly low compared to the Galactic halo field.   

Regarding possible systematic errors, the referee reminds us that the CNO abundances derived from molecular 
features depend sensitively on the $T-{\tau}$ structure of the upper layers of the model atmospheres.  Computations 
of 3D hydrodynamic model atmospheres suggests that the adequacy of the $T-{\tau}$ structure of 1d hydrostatic model atmospheres declines with metallicity \citep[compare, e.g., Figures 1 and 2 of][]{Asplund2005}. Any metallicity-dependent component of this difference would not be removed by a differential analysis with respect to the Sun.  One way to mitigate 
the impact of such possible effects is to compare our abundances with those in similar metallicity Galactic halo stars 
that are also derived from molecular features; we do this in \S3.2.

\section{RESULTS AND DISCUSSION}

We group and list 22 elements with abundance determinations in HD 134439/440 in the first two 
columns of Table \ref{tab:abundances}.  Columns 3 and 4 give the [X/Fe] values for HD 134439 and 
134440, respectively; CNO and Ag abundances are those derived in this study, Fe is taken from
\citet{1997AJ....113.2302K}, and the remainder are from \citet{2006MNRAS.370.2091C}.  Column 5 contains 
the average [X/Fe] and estimated uncertainty.   Condensation temperatures for the elements 
\citep[from Table 8 of][]{2003ApJ...591.1220L} are listed in the final column. 

\subsection{Abundances and T$_{\rm C}$}
\begin{figure}
\plotone{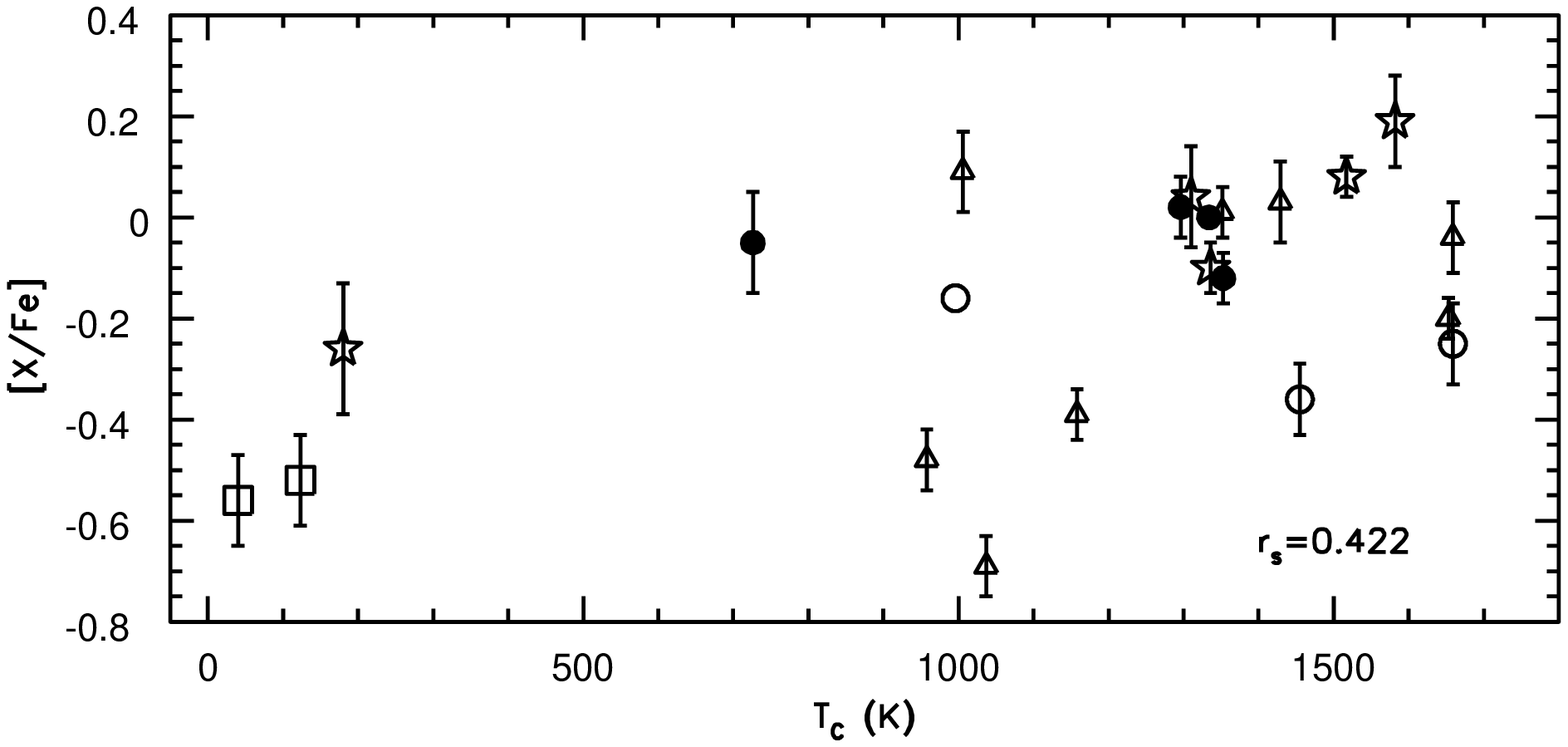}
\caption[{[X/Fe]} v.s. T$_{\rm c}$ for HD 134439 and HD 134440]
{Mean [X/Fe] of HD 134439 and HD 134440 for the 22 elements 
in Table \ref{tab:abundances} versus condensation temperature $T_{\rm C}$. 
The open star symbols indicate $\alpha$ elements, triangles designate odd Fe group 
elements, filled circles are for even iron group elements, open circles are for $n$-capture 
elements, and open squares are C and N.  
One $\sigma$ error bars for the mean abundance ratios are shown. 
The Spearman rank correlation coefficient for the data is 0.422.}
\label{fig:xfe}
\end{figure}

We plot the average HD 134439/440 abundance ratio versus condensation temperature in Figure \ref{fig:xfe}.
The CNO abundance ratios (at low T$_{\rm c}$) are comparatively lower than those of the majority of 
refractory elements at high T$_{\rm c}$.  The size of the Spearman's rank coefficient (0.422), which measures 
the correlation between the abundances and T$_{\rm c}$ via the monotonicity 
of the rankings of the two variables with no underlying assumptions concerning their parent distributions 
\citep{WJ2003}, indicates only a 5\% probability that the observed correlation could occur by chance in
an uncorrelated sample--not inconsistent with claims of chemical pollution via planetesimal/refractory accretion 
\citep{2003ApJ...598L..47S}.  

The mean difference in abundance ratios (HD 134439 $-$ HD 134440;) from Table \ref{tab:abundances} is 
$0.02$ dex with scatter ($\pm 0.08$ dex) that is consistent with uncertainties.  Given  a) the two stars' 
convection zone depths that differ by ${\ge}15\%$ based upon our T$_{\rm eff}$ and the model estimates from Figure 2 of 
\citet{2002ApJ...580.1100R}, and b) possible differences in the mass/composition of material putatively 
accreted by each star, remarkable fine-tuning would have been required for them to now exhibit
indistinguishable post-accretion abundance patterns.  We are thus led to examine other interpretations of 
the [X/Fe] verus T$_{\rm c}$ relationship.  

\subsection{Comparisons with the Galactic Halo}
One such interpretation is that the [X/Fe]-T$_{\rm c}$ correlation is simply an unremarkable result of 
Galactic chemical evolution.  We explore this possibility by comparing HD 134439 and 134440 to Galactic 
halo stars of similar [Fe/H], adopting typical [X/Fe] values of halo stars with [Fe/H]$\sim -1.50$ using the 
observational data from \citet{2006ApJ...653.1145K}, \citet{Zhang2009}, and \citet{NS10} (for their high-${\alpha}$ stars) 
with a few exceptions.  First, we believe the high [N/Fe] values at [Fe/H]$\sim-1.50$ shown in 
\citet{2006ApJ...653.1145K} are due to mixing in evolved giants.  The near-UV NH-based [N/Fe] 
values in unmixed or unevolved stars from the studies of \citet{2005A&A...430..655S} and 
\citet{2004A&A...421..649I} consistently indicate that [N/Fe]=$0.0$ at [Fe/H]$=-1.5$.  The 
primary nucleosynthetic behavior of N this result implies has long been suspected--grave
uncertainties in previous N abundance determinations not withstanding (e.g., 
\citealt{1989ARA&A..27..279W} and \citealt{2003ASPC..304..361P}).  A value of [N/Fe]$=0.0$ at
[Fe/H]$=-1.5$ is also in excellent agreement with a metal-poor extension of the 
CN-based [N/Fe] ratios in more metal-rich giants from \citet{2000A&A...356..238C} that are 
presumably immune from the effects of CN-cycling.  

Second, we adopt the OH-based oxygen ratios in metal-poor field stars of  \citet{1999Boes,2011Boes} to 
compare with our molecular-based results in HD 134439/440.  Third, we adopt the CH-based carbon ratios 
in metal-poor field stars from \citet{Tomkin92}, \citet{Gratton00}, \citet{Simmerer04}, and \citet{Lai07} to 
compare with our molecular-based 
results.    Fourth, we adopt Al and K data compiled by \citet[and references therein]{1995ApJS...98..617T}.  Fifth, 
we adopt the Galactic Y and Ba data compiled by \citet{2004ApJ...601..864T}.  Finally, Galactic Ag data is taken 
from \citet{1998AJ....116.2489C}. 

\begin{figure}
\plotone{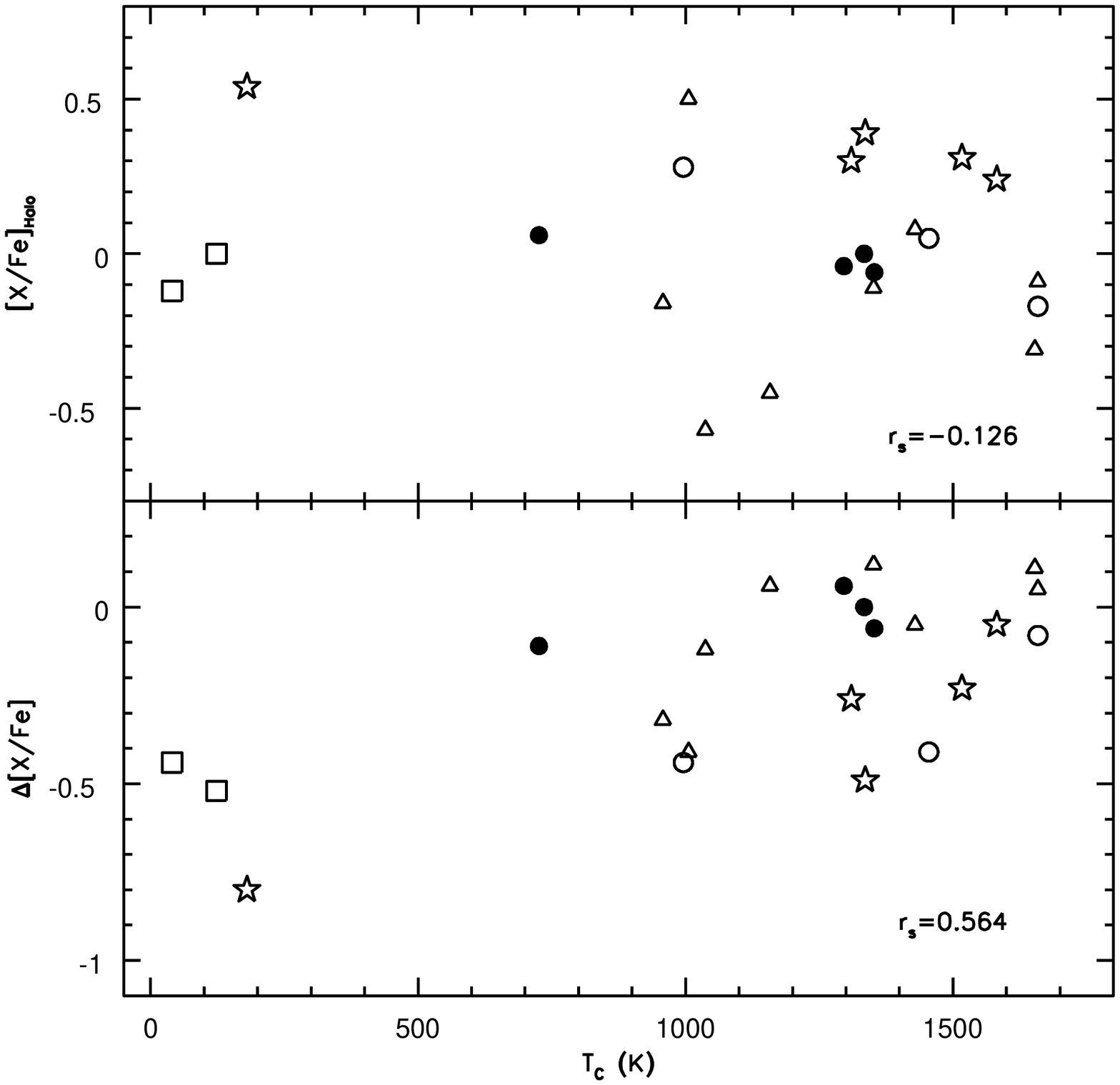}
\caption[{[X/Fe]} versus $T_{\rm C}$ for Galactic halo field stars]
{Mean [X/Fe] for Galactic halo field stars at [Fe/H]${\sim}-1.50$ (see text)  
are plotted against $T_{\rm C}$ in the upper panel.  The lower panel shows the differences 
between mean galactic halo [X/Fe] and values of HD 134439 and 134440 (in the 
sense our stars minus halo stars).  The symbols are as in Figure \ref{fig:xfe}.} 
\label{fig:halo}
\end{figure}

The mean Galactic halo abundance ratios at [Fe/H]${\sim}-1.5$ have a negligible correlation with 
$T_{\rm C}$ (upper panel of Figure \ref{fig:halo}).  The contrast between the mean Galactic halo
field and HD 134439/134440 is seen in the lower panel of Figure \ref{fig:halo}; the differential 
abundance ratios  (our stars $-$ halo) demonstrate a correlation with $T_{\rm C}$ at the ${\sim}99.5$\% 
confidence level.  This is largely driven by CNO: the Spearman-based correlation excluding CNO 
is significant at only the ${\sim}85\%$ confidence level.  

Given the multiplicity of nucleosynthetic sites for CNO production and uncertainties in our understanding 
of non-standard stellar mixing, winds, etc, it is possible that the [X/Fe]-$T_{\rm C}$ correlation in HD 
134439/440 is simply the consequence of integrated stochastic effects of their chemical evolution.  
However, the unusual $V_{\rm LSR}$, $R_{\rm apo}$, and $Z_{\rm max}$ values of our 2 stars suggest 
their relevant preceding chemical evolution history did not transpire in the Galaxy \citep{CLL96}.  Thus, 
we explore alternate explanations of the abundance correlation and the low-$\alpha$ 
phenomenon.  

\subsubsection{O, Mg, Al, and Proton-Capture}
There are interesting similarities between the HD 134439/440 abundance patterns and those of Galactic 
globular cluster red giants, for which considerable evidence of proton-capture nucleosynthesis has 
been observed \citep{1994PASP..106..553K, 1996AJ....112..545P, 1997AJ....113..279K, 1998AJ....115.1500K}. 
Deep mixing--in situ or in previous stellar generations having enriched the cluster gas--has 
apparently enhanced some cluster giant N, Na, and Al abundances by conversion of C O, Ne, and Mg 
nuclei through proton-capture.  As a result, one can observe low [O/Fe] and [Mg/Al] ratios for many 
cluster giants compared to enhanced ratios observed in halo field giants and  those expected from Type II 
SNe nucleosynthesis.  The mean Galactic halo field data (Figure \ref{fig:halo}) evince [O/Fe]${\sim}0.57$ and 
[Mg/Al]${\sim}0.70$ at [Fe/H]${\sim}-1.50$.  The corresponding ratios for HD 134439/440 
(Table \ref{tab:abundances}) are 0.83 and 0.60 dex lower, and more similar to the ratios exhibited by many 
globular cluster giants.  

The Na and N abundances of HD 134439/440, however, exclude proton-capture as the origin of this
similarity.  Cluster giants with 
[O/Fe]${\sim}-0.25$ (like HD 134439/440) demonstrate [Na/Fe]${\sim}0.40$ (see Figure 4 of 
\citealt{1998AJ....115.1500K}); the mean [Na/Fe]$\sim -0.48$ for HD 134439/440 is not similarly enhanced.  
In fact, the HD 134439/440 Na abundance is low compared to the mean halo field data ([Na/Fe]${\sim}-0.16$).  
Similarly, our N abundance is also lower than the proton-capture processed material in cluster giants or in the 
mean halo field.  The deep mixing models shown in Figures 1-3 of \citet{LH95} conserve the sum of C$+$N$+$O; 
our Figure 5 indicates that this sum in HD 134439/134440 is ${\le}20$\% (by number) of the sum for halo field 
stars at [Fe/H]${\sim}-1.5$.  Given these key comparisons, we reject proton-processing as the agent responsible 
for our stars' anomalous abundances with respect to the halo field. 

\subsection{Extragalactic Origins: Dust}

Because Galactic halo field stars showing low [$\alpha$/Fe] and [Na/Fe] in the studies of  
\citet{1997A&A...326..751N} and \citet{NS10} demonstrate distinctive kinematics, these authors hypothesize 
an extragalactic origin (e.g., dwarf galaxies) for such chemically anomalous stars.  HD 134439/440 
also exhibit low [$\alpha$/Fe] and [Na/Fe] as well as large R$_{\rm max}$ ($\sim$ 43 kpc), and a
dSph origin has previously been posited for these two stars \citep{1997AJ....113.2302K, 2006MNRAS.370.2091C}.  
The latter authors suggest formation in a high dust-to-gas environment, which might explain the HD 134439/440 
abundance ratio-$T_{\rm C}$ correlation and low $\alpha$-element abundance ratio, as an alternative to the 
planetesimal accretion mechanism proposed to explain these abundance data \citep{2003ApJ...598L..47S}. 
In the \citet{2006MNRAS.370.2091C} scenario, dust from ancestral stellar generations 
is enriched in (high $T_{\rm c}$) refractory elements relative to (low $T_{\rm c}$) 
volatiles--consistent with the opposite pattern evinced by gas-phase ISM results \citep{Field74};  
this dusty pollutant, combined with natal pre-stellar material, would end up back in the 
HD 134439/134440 photosphere we observe today, creating a correlation of abundance
with $T_{\rm c}$.  

\begin{figure}
\plotone{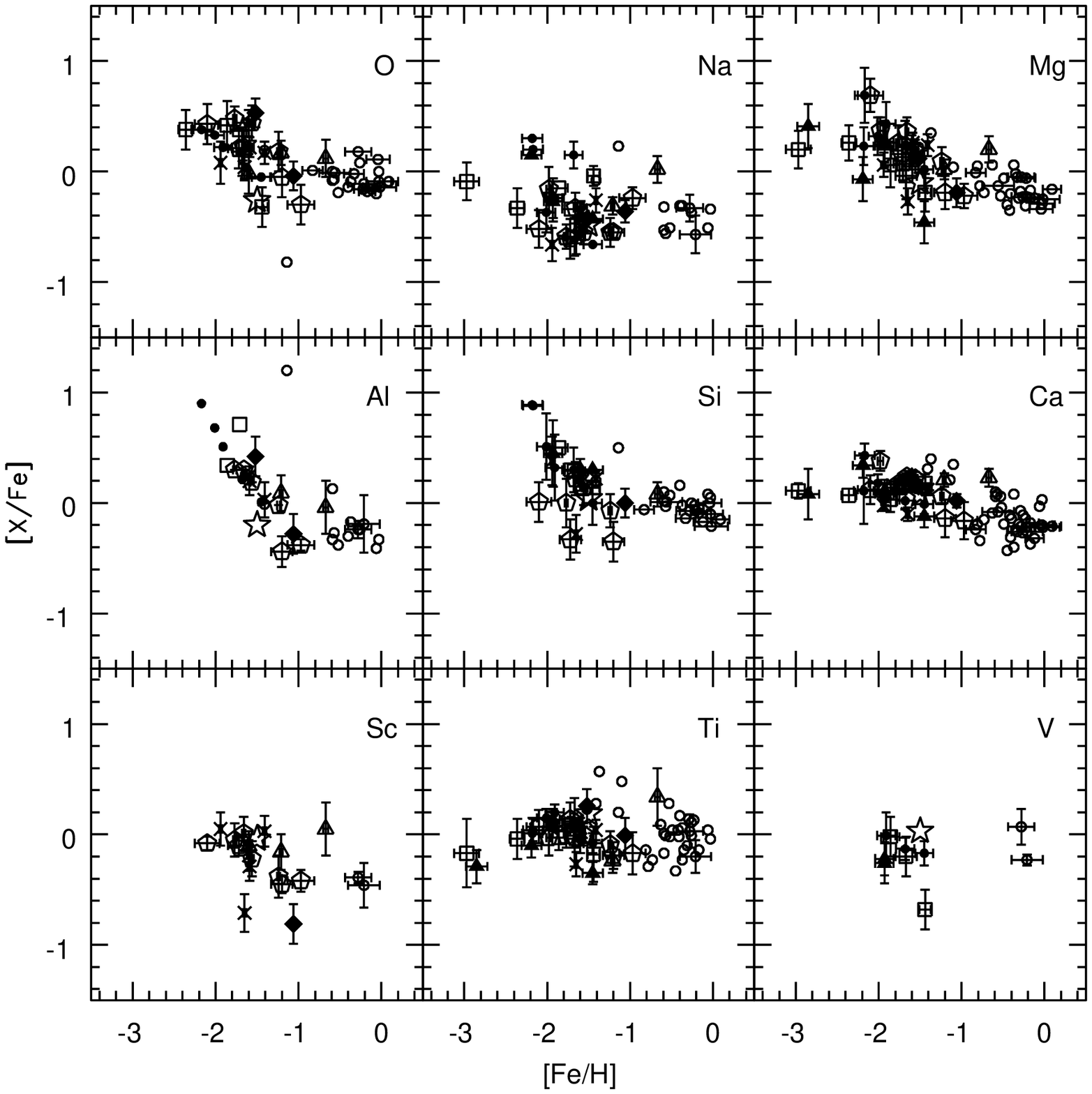}
\caption[{[X/Fe]} for dSph Stars (O-V)]
{[X/Fe] as a function [Fe/H] for stars in dSph galaxies from various studies (see
text).  The dSph galaxies are shown in different symbols: open circles for Sgr, open pentagons 
for Scl, open squares for Dra, open triangles for For, filled circles for UMi, filled triangles 
for Sex, filled diamonds for Leo, and crosses for Car.  The open star symbol indicates the mean 
[X/Fe] for HD 134439 and HD 134440.}
\label{fig:allfe1}
\end{figure}

We show the HD 134439/440 abundance ratios and those for dSph red giants in Figures \ref{fig:allfe1} 
and \ref{fig:allfe2}.  The abundances for Sagittarius (Sgr) are obtained from 
\citet{2000A&A...359..663B, 2004A&A...414..503B}, \citet{2005A&A...441..141M}, and 
\citet{1999ASPC..192..150S}.  We use results from \citet{2001ApJ...548..592S, 2003AJ....125..684S} 
for giants in Draco (Dra), Ursa Minor (UMi), Sextans (Sex), Sculptor (Scl), Fornax (For), 
Carina (Car), and Leo I.  We have also adopted the \citet{2005AJ....129.1428G} abundance measurements 
for four red giants in Scl.  Due to the lack or sparseness of dSph C, N, Eu, and Ag measurements (particularly at
[Fe/H]${\sim}-1.5$), panels for these elements 
are omitted.  A datapoint corresponding to the average HD 134439/440 abundance ratio is included in each 
panel of Figures 6 and 7, but is not easily discerned within the dense dSph scatter.  Indeed, the HD 134439/440 
abundance ratios fit within the manifest continuum of in situ dSph abundances that result from their individual 
histories \citep{2008A&A...481..635L}.  

Direct evidence against an unusually dusty pre-stellar environment for HD 134439/440 is, however, 
found in their Be abundances.  We determined these via synthesis of the ${\lambda}3130.4,3131.1$ \ion{Be}{2} 
doublet as accomplished for halo field stars in \citet{2011Boes}.  While the ionized doublet of this trace element 
in our metal-poor high-gravity stars is vanishingly weak, we set an upper limit of log $N$(Be)$=-0.7$ (on the 
usual scale where log $N$(H)$=12.$) in HD 134439.   This upper limit is ${\sim}0.7$ dex lower than the abundance 
of Galactic halo field stars of similar [Fe/H] seen in Figure 9 of \citet{2011Boes}.  A deficiency of this 
high-$T_{\rm C}$ (1450 K) element is inconsistent with the enhancement expected if our star was formed in a 
refractory-rich environment or had accreted refractory-rich material.  
 
Instead, the upper limit is consistent with the Be abundances of Galactic halo field stars of 
similar [O/H] (Figure 11 of Boesgaard et al.~2011). That the HD 134439 Be abundance appears anomalously 
low relative to Galactic halo field stars when referenced to Fe, but unremarkable when referenced to O is 
qualitatively in accord with: {\ }a) the spallation origin of Be \citep{1970Natur.226..727R} linking its nucleosynthetic 
history to that of O, and {\ }b) the well-known differences between [O/Fe] (and other [$\alpha$/Fe] ratios) in the 
Galactic halo and dSphs.  For example, Figure 8 from \citet{2000AJ....120.1056K} and our Figure 6 indicates 
that [O/Fe] is ${\ge}0.4$ dex higher at [Fe/H]$=-1.5$ in the Galactic halo; this difference is confirmed by the Galactic 
and dSph [Mg/Fe] ratios seen in, e.g., Figure 1 of \citet{2006A&A...447...81T} or Figure 11 of \citet{Tolstoy2009}

The referee notes that Galactic halo field Be abundances are determined in stars ${\sim}500-1000$ K warmer
than HD 134439.  This raises the possibility that the initial Be abundance of HD 134439 might have been larger than that in 
hotter Galactic field stars of similar [O/H], but now is lower due to greater stellar depletion in our cooler dwarf.  Unfortunately, 
the Be-$T_{\rm eff}$ trend in metal-poor stars that would trace the effects of differential depletion is not observationally defined.
However, stellar models suggest that the difference in Be depletion in 5000 and 6000 K old metal-poor dwarfs is zero or in the
opposite sense suggested (i.e., hotter model stars deplete slightly more Be).  The $Z=0.001$ standard stellar model results
in Table 3A of \citet{PDD} suggest no depletion in old 5000 K or 6000 K stars.  The rotational stellar models in Tables 4A-4C of 
\citet{PDD} suggest that Be depletion may be ${\le}0.2$ dex {\it greater} at 6000 K than at 5000 K.  Such a modest effect, if real, 
would only strengthen our conclusion about the difference in Be abundance between HD 134439 and the Galactic halo field. 

We note that there is no evidence from the ${\lambda}3131$ features of 
any Be in the cooler, less massive HD 134440.  A real difference in our two stars' Be content may not be surprising.
The temperature of the surface convection zone base, its temporal evolution, and thus the extent and 
length of Be destruction due to pre-main-sequence and main-sequence convective mixing, are highly 
sensitive functions of mass in the mass regime occupied by our stars \citep{1966ApJ...144..103B}. 

\subsection{Stochastic Nucleosynthetic Origins: Type II Contributions} 

None of the preceding mechanisms--Galactic halo-like chemical evolution, (proto-)planetary accretion, 
formation in an unusually dusty environment, or formation from material enriched in stellar $p$-capture 
products--provides a plausible or convincing explanation of the [X/Fe]-$T_{\rm C}$ relation in HD 134439/440.   
We thus consider whether this relation is simply an illusory one stemming from these stars' 
unique chemical evolutionary history.  Motivation and guidance in doing so is provided by \citet{2003AJ....125..707T}, 
who analyze stellar abundances in four dSph galaxies and find $\alpha$-element abundance ratios suggesting 
negligible nucleosynthetic contributions from stars of progenitor mass 
${\ge}15-20$ M$_{\odot}$.  \citet{2003AJ....125..707T} note that such a mass function truncation may be only 
effective, and other possible explanations (e.g., star formation efficiency and metal-loss by winds) having the 
same observable consequence do exist .   We explore the effects of mass function truncation here 
using the Type II yields of \citet{2006ApJ...653.1145K}, and find that explosive nucleosynthesis in Type II 
SNe of low progenitor mass provides an attractive explanation for the abundance patterns of 
HD 134439/440. 

The [O/Fe] ratio and those of the higher $T_{\rm C}$ $\alpha$-elements Mg, Si, and Ca are lower in HD 134439/440 
than in the mean Galactic halo (columns 5 and 6 of Table 12).    The Galactic potassium abundance ratios used 
in Figure \ref{fig:halo} exhibit behavior strongly reminiscent of traditional $\alpha$-elements; the markedly low
[K/Fe] ratio of our stars relative to those in the Galactic halo at similar [Fe/H] (a difference of 0.4 dex as seen in Table 
12) is thus a notable similarity.    In sum, the HD 134439/440 $\alpha$-element ratio resembles the sub-Galactic halo 
values seen in dSph stars.   We now seek additional evidence in the HD 134439/440 abundance patterns that might 
support the idea of their origin in a dSph-like environment whose prior chemical evolution was dominated by the 
effects of a small number of discrete nucleosynthetic events characterized by mass function truncation. 

Such supporting evidence can be found in these stars' intra-$\alpha$ element ratios.  In particular, [O/Fe] is 0.15-0.45 
dex lower than ratios of other $\alpha$-elements; we suggest this has little to do with $T_{\rm c}$-related 
behavior per se.  Instead, we note that the [O/Fe] yields in, e.g., $Z=0$ Type II models of \citet{2006ApJ...653.1145K} 
are steep functions of mass--some 0.60 dex lower for progenitors of 14 M$_{\odot}$ than for 20 M$_{\odot}$.  
In contrast, the [Mg/Fe] and [Si/Fe] yields are much shallower functions of mass; their yields change by 
$\le 0.20$ dex over the same progenitor mass range.  Formation from material enriched by Type II SNe biased 
towards low progenitor mass is qualitatively consistent with low [O/Fe] compared to [Mg,Si/Fe] in HD 134439/440.
Indeed, for progenitor mass of 14 M$_{\odot}$, the [Mg,Si/Fe] yields are some 0.30 dex larger than for [O/Fe]--a 
difference in excellent agreement with observed values. 

The HD 134439/440 [Na/Fe] ratio is anomalously low as well, lying some 0.3 dex below the Galactic halo mean
(Table 12).  As is the case for [O/Fe], the \citet{2006ApJ...653.1145K} model [Na/Fe] yields also drop precipitously 
over the 20 to 14 M$_{\odot}$ progenitor mass range.   In Figure 8, we compare the observed abundance ratios 
in HD 134439/440 with those of the \citet{2006ApJ...653.1145K} yields for a 14 M$_{\odot}$ progenitor.  The observed
[Na/Fe] is well-matched by the low mass progenitor yield.   The distinctly sub-solar ratio [Na/Fe] ratio (${\sim}-0.50$)
of HD 134439/440 is thus easily explained by our proposed nucleosynthetic origin without recourse to $T_{\rm c}$-related 
effects. 

The [C/Fe] and [N/Fe] values of HD 134439/134440 are also low compared to the Galactic halo (Table 12); the 
differences are even a factor of two lower than for [O/Fe].   This too needs have little to do with $T_{\rm c}$-related 
effects.  Chemical evolution models of dSph galaxies predict [N/O] and [C/O] ratios as low as $-1$ (Figure 13 of 
\citealt{2008arXiv0802.1203C}), which easily accommodates the values of $\sim -0.30$ in our stars if they formed 
in such an environment.    

\begin{figure}
\plotone{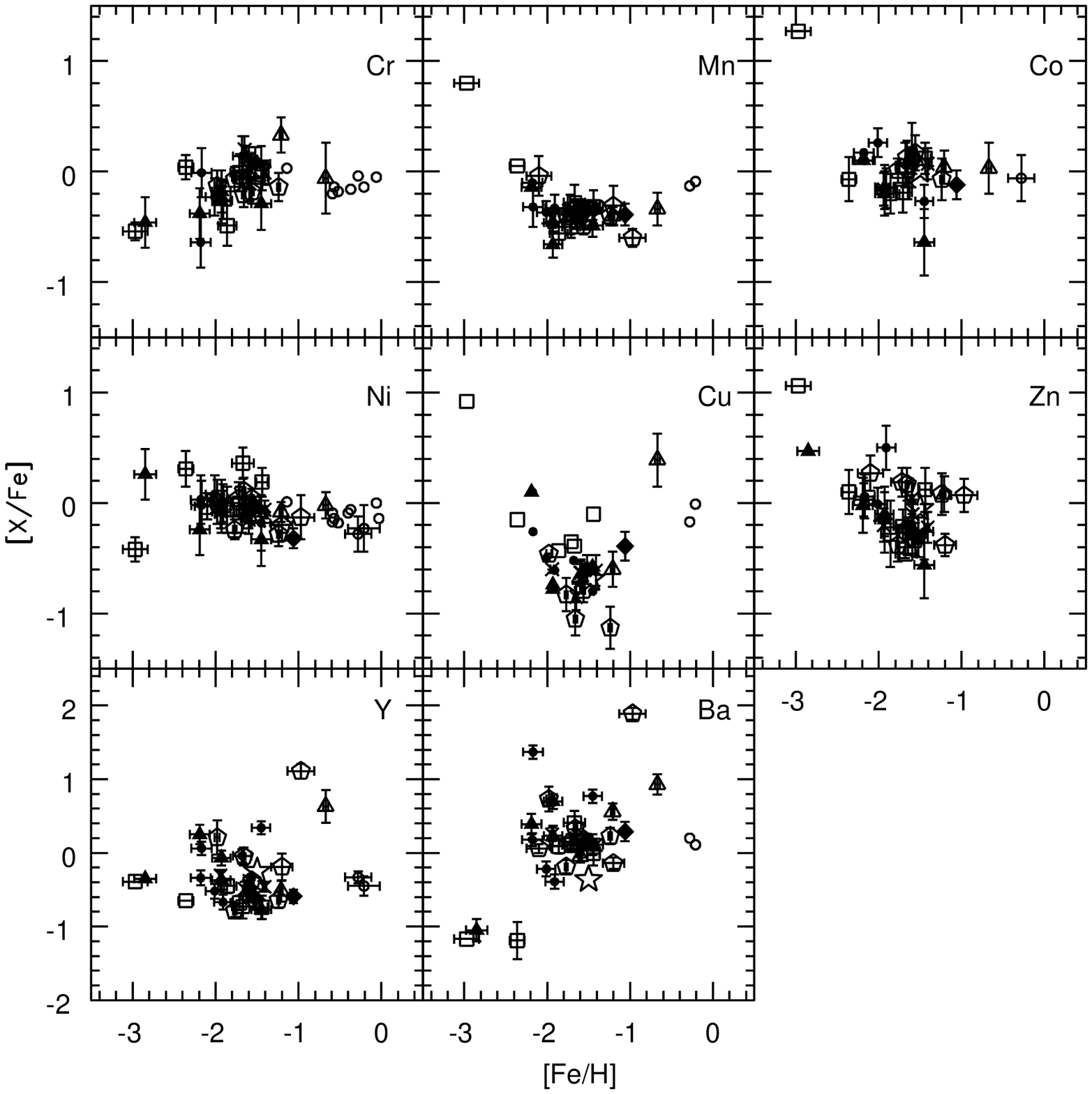}
\caption[{[X/Fe]} for dSph Stars (Cr-Ba)]
{Same as Figure \ref{fig:allfe1} for additional element ratios.}
\label{fig:allfe2}
\end{figure}

\begin{figure}
\plotone{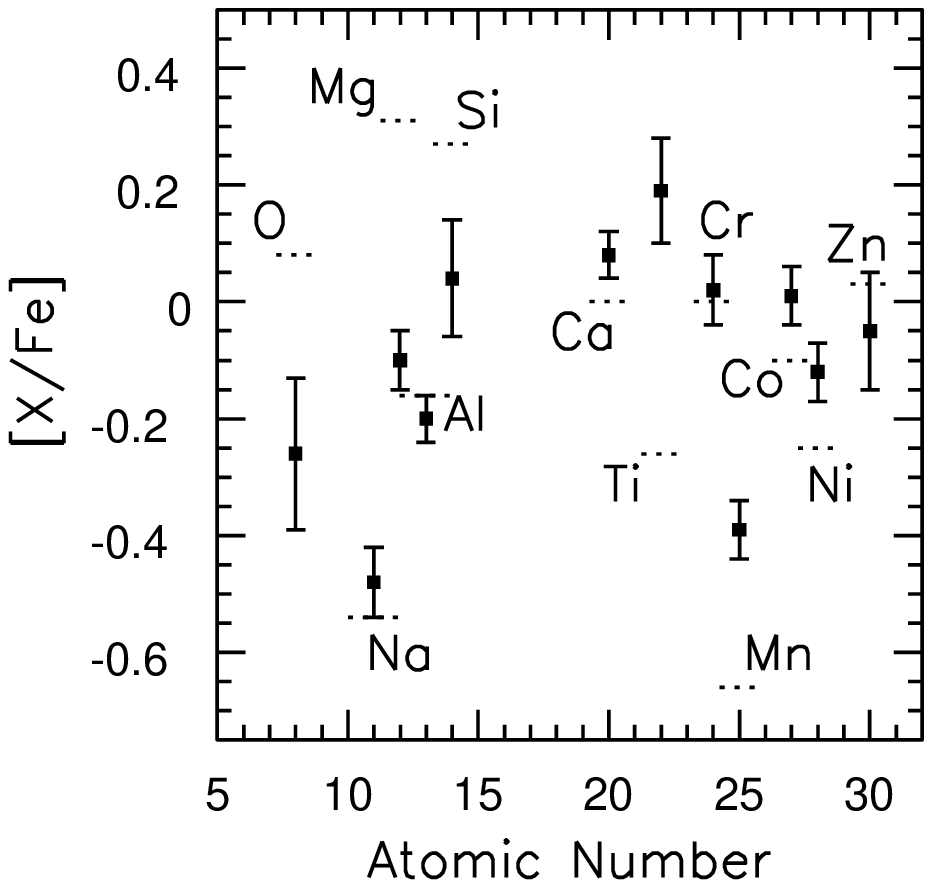}
\caption[X/Fe versus Kobayashi et al. Z=0 M=14Sun yields]
{Observed mean [X/Fe] ratios for HD 134439/134440 (data points) compared to the abundance ratios from the 
$Z=0$, $M=14$ M$_{\odot}$ progenitor models of Kobayashi et al. (2006; horizontal dashed lines.} 
\label{fig:obsmod}
\end{figure}

Figure \ref{fig:obsmod} shows that the [Zn,Ni,Co/Fe] values of our stars agree with the ${\sim}14$ M$_{\odot}$ 
progenitor mass yields of \citeauthor{2006ApJ...653.1145K}. The observed [Al,Cr/Fe] ratios are seen to be 
virtually identical to these 14 M$_{\odot}$ progenitor yields. The value [Cu/Fe]${\sim}-0.70$ for our 2 stars also 
need not be associated with T$_{\rm C}$-related effects inasmuch as it is not significantly different 
than the ratio observed in Galactic halo or dSph stars (Figures \ref{fig:halo} and \ref{fig:allfe2}).  
As noted by \citet{1995ApJS...98..617T}, Galactic halo Cu data are consistent with the Type II yields 
of \citet{1995ApJS..101..181W}, suggesting Cu production is dominated by explosive nucleosynthesis 
and not production in Type Ia supernovae or low-to-intermediate mass stars. 

A couple puzzles remain. First, the observed [Mn/Fe] value is 0.30 dex larger than the 
\citet{2006ApJ...653.1145K} [Mn/Fe] yield at 14 M$_{\odot}$; however, it's interesting to note that 
the observed value is  not anomalous with respect to dSph stars in this regard (Figure \ref{fig:allfe2}). 
Second, as already noted by \citet{2006MNRAS.370.2091C}, our stars' [Ti,Ca/Fe] values are both 
{\it larger\/} than [Mg/Fe], which is consistent neither with the \citet{2006ApJ...653.1145K} [X/Fe] 
yields at low progenitor mass nor the dSph results of \citet{2004AJ....128.1177V} who find [Ti,Ca/Fe] 
ratios smaller than [Mg/Fe].  We return to the [Ti,Ca/Fe] ratios in \S 3.6. 

\subsection{Type Ia Contributions} 
The preceding comparisons suggest that the abundance ratios of HD 134439/440 are dominated by
nucleosynethic products of low mass progenitor Type II SNe.  We find no observable evidence of Type Ia contributions 
in the abundance data.  Such contributions 
are of possible interest in explaining the absolute levels of [O/Fe] and [Mn/Fe], which 
are some 0.30 dex lower and higher (respectively) than the \citet{2006ApJ...653.1145K} 14 M$_{\odot}$ 
progenitor yields; these differences are qualitatively consistent with lower and higher 
(respectively) ratios expected from Type Ia event yields.   

We constructed simple Type II/Ia mixture models from the \citet{2006ApJ...653.1145K} 
Type II yields for a given progenitor mass and the Type Ia yields from 
\citet{2003NuPhA.718..139T} as tabulated in \citet{2004A&A...425.1029T}.  The Type Ia yield was 
weighted so as to reproduce the observed mean [O/Fe] of our stars when combined with the Type II 
yields.  Resulting [X/Fe] yields from the Type II/Ia mixture were then computed.  This process was then repeated 
using different Type II progenitor masses to produce multiple Type II/Ia mixture models.  All such mixtures are unable to 
consistently reproduce the observed abundance ratios.  We typically find that, simultaneously, the 
[Ni/Fe] yields are ${\ge}0.30$ dex larger than observed, [Mn/Fe] becomes overenhanced by 
${\ge}0.25$ dex, [Na/Fe] is ${\ge}0.30$ lower than observed, and the predicted [Ti/Fe] yield 
ratios become even lower compared to the data.

\subsection{\emph{Neutron-capture Elements}}

\subsubsection{Heavy $n$-capture elements: Ag, Eu, and Ba} 

\begin{figure}
\plotone{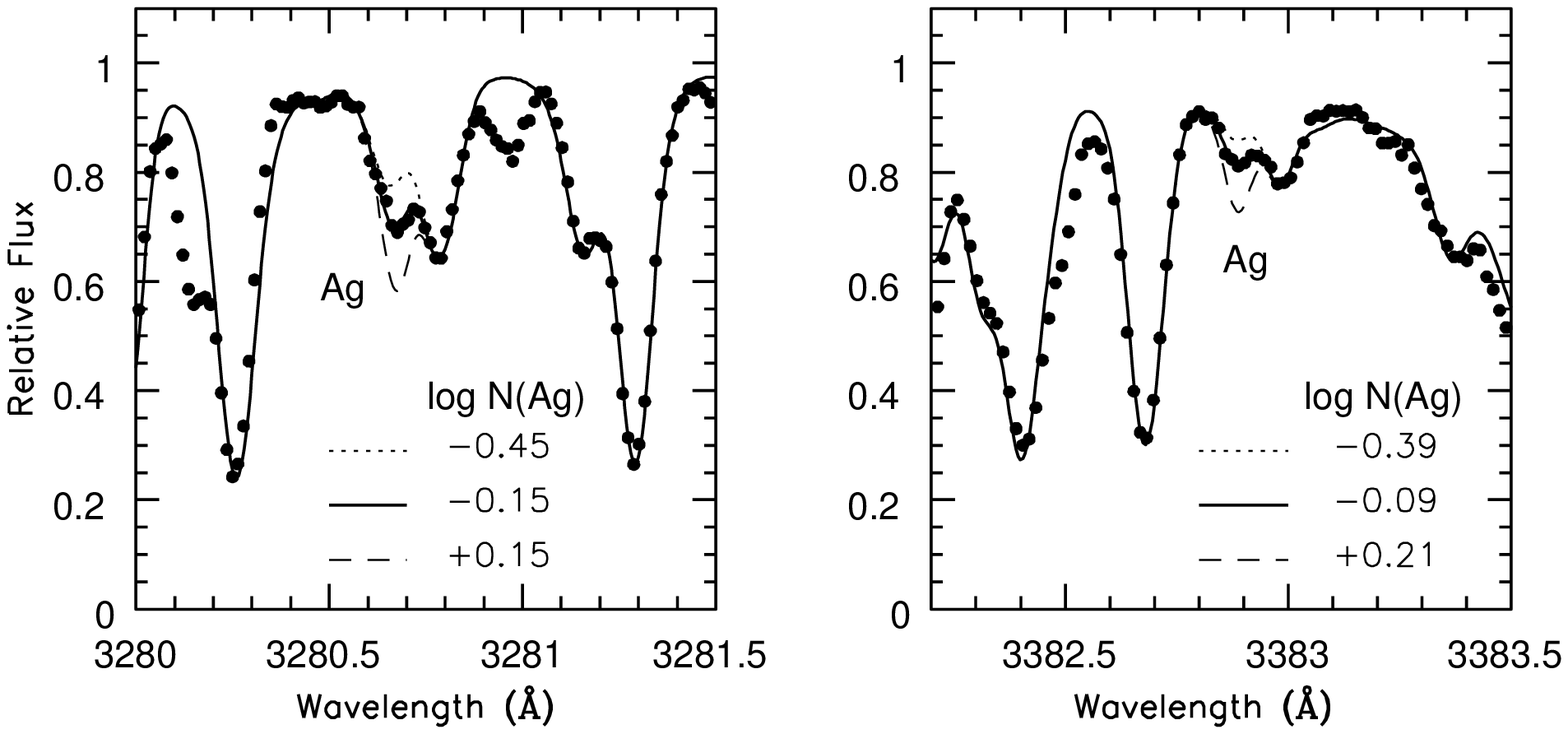}
\caption[Silver Syntheses for HD 134439]
{Syntheses of \ion{Ag}{1} features at 3280{\AA} (left panel) and 3382{\AA} (right panel) for HD 134439.
Filled circles indicate observed spectra. The solid line is the best fit Ag abundance; dotted and dashed 
lines are $\pm$0.30 dex deviations from the best fit abundance.}
\label{fig:ag39}
\end{figure}

While silver may seem an unconventional choice of element with which to explore $n$-capture 
nucleosynthesis, its great utility comes from observably strong transitions of {\it neutral} 
features in the near-UV.  The observed $n$-capture transitions upon which most of our knowledge 
of Galactic and dSph abundance patterns are built are 
primarily singly ionized lines in low gravity red giants that can be vanishingly weak in our metal-poor cool dwarfs.  
We measured Ag abundances for HD 134439 and HD 134440 from synthesis of the ${\lambda}$3280 and 
${\lambda}$3382 \ion{Ag}{1} line regions.  Line lists were compiled for the Ag line regions using 
VALD and assuming a solar Ag abundance of $\log N({\rm Ag})_{\odot}=1.54$.  We find [Ag/Fe]$=-0.19$
and $-0.13$ for HD 134439 and 134440 (Figure \ref{fig:ag39}) with an uncertainty of 0.14 dex.  
\citet{1998AJ....116.2489C} determined [Ag/Fe] in the metal-poor cool halo dwarf HD 103095, which 
has physical properties (T$_{\rm eff}=5007$ K, $\log g=4.65$, [Fe/H]$=-1.27$) similar to HD 134439, 
obtaining [Ag/Fe]$=+0.28{\pm}0.17$ dex from the same \ion{Ag}{1} lines.   The 
significantly lower [Ag/Fe] value compared to that in HD 103095 can easily be seen by comparing 
the relative strengths of the neighboring \ion{Ag}{1} and \ion{Fe}{1} lines in Figure \ref{fig:ag39} 
with those of Figure 3 in \citet{1998AJ....116.2489C}.

\begin{figure}
\plotone{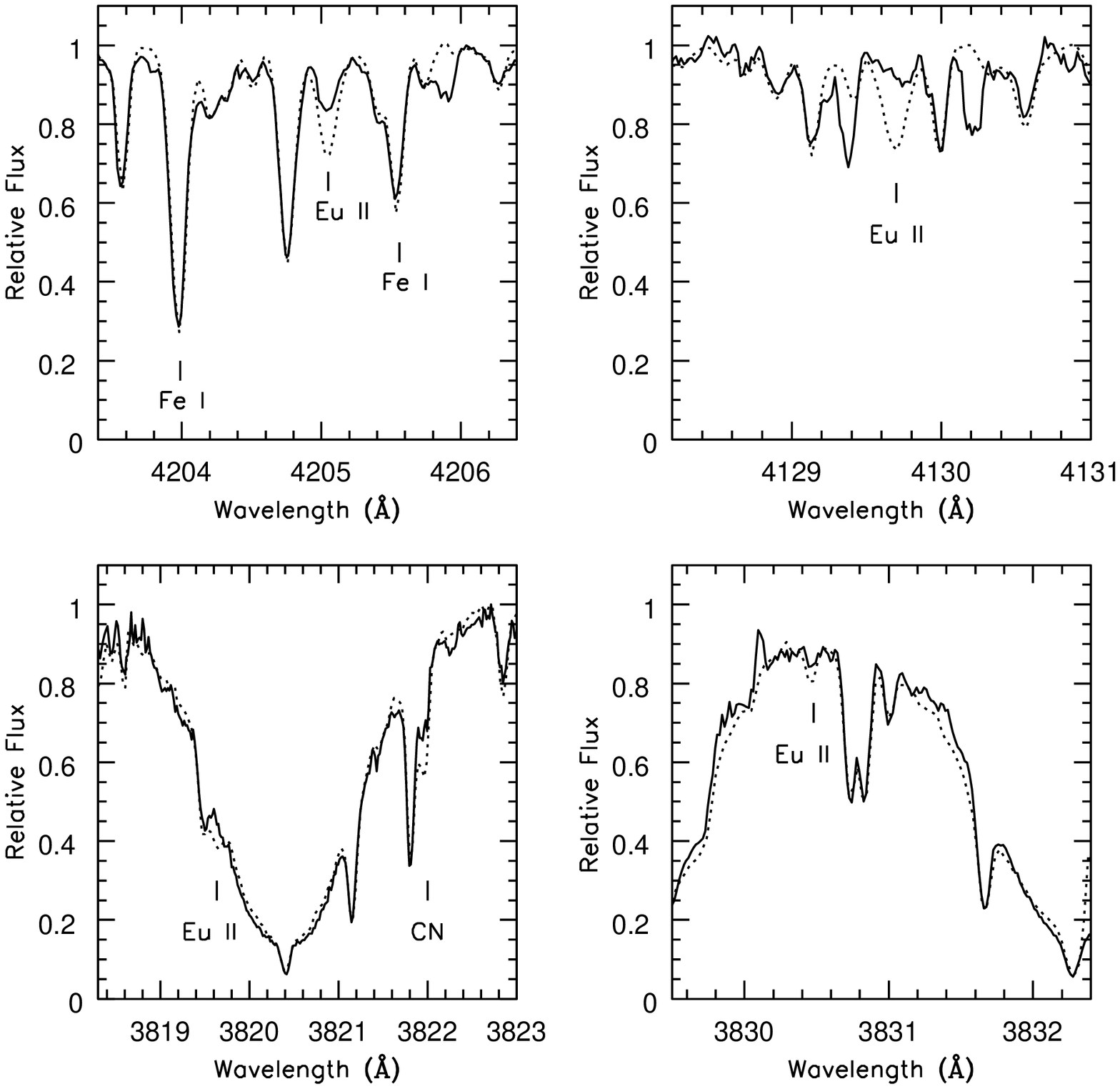}
\caption[Europium Features]
{A comparison of 4 different \ion{Eu}{2} features in consistently normalized observed spectra of HD 134439 
(solid line) and HD 103095 (dotted line).  The Eu lines in HD 103095 are consistently stronger relative to 
neighboring metal lines compared to HD 134439.  The feature near 4130.2{\AA} is believed to be an artifact in 
the HD 134439 spectrum.  The feature near 4129.4{\AA}, however, appears real; \ion{Yb}{2} lines are in present 
in the VALD list at this wavelength region, but we do not observe what should be stronger known \ion{Yb}{2} 
features at other wavelengths.}
\label{fig:eu}
\end{figure}

We sought to compare [Eu/Fe] in HD 134439/134440 and 103095 by synthesizing the $\lambda$4205 
\ion{Eu}{2} lines.  For HD 103095, we utilized a high S/N, R$\sim$60,000 spectrum obtained with 
the McDonald Observatory 2.7m 2dCoude spectrograph and the stellar parameters adopted by 
\citet{1997PASP..109..776K}: T$_{\rm eff}=5050$ K, [Fe/H]$=-1.31$, $\log g=4.5$, and $\xi=1.5$ 
km/s; the McDonald spectrum does not allow for a direct comparison of Ag abundances because it 
does not extend to the atmospheric cutoff region near 3300 {\AA}.  The line profile at 4205.04{\AA} 
appears blended in our stars and we encountered difficulties simultaneously and consistently 
fitting the wings of the line in the stars and the Sun (for which we assumed $\log N({\rm Eu})_{\odot}=0.51$
in making $gf$-value adjustments).  At present, we are only confident in establishing upper 
limits on [Eu/Fe] by fitting the ${\lambda}4205.04$ line depth.  This yields ratios of 
[Eu/Fe]${\lesssim}+0.31$ for HD 134439, ${\lesssim}+0.42$ for HD 134440, and ${\lesssim}+0.70$ for 
HD 103095.  The suggestion that [Eu/Fe] is ${\sim}$0.30-0.40 dex larger in HD 103095 is confirmed in 
Figure \ref{fig:eu}, where we compare the observed spectra of HD 103095 (dotted line) with 
HD 134439 (solid line) over four spectral regions containing \ion{Eu}{2} features; these features 
are consistently stronger relative to other metal features in HD 103095 compared to HD 134439. 

Our Eu and Ag results and the Ba abundance given in Table 12 indicate that [Ba, Ag, Eu/Fe] ratios 
are consistently lower in HD 134439/134440 by ${\sim}0.40$ dex compared to the mean Galactic halo 
field at [Fe/H]$=-1.5$. Marked deficiencies (${\ge}1$ dex) in the heavy $n$-capture elements are 
seen in the low-$\alpha$ star BD$+$80 245 ([Fe/H]${\sim}-2.05$), which has [Ba/Fe]$=-1.87$ and 
$-1.89$ and [Eu/Fe]$=-1.04$ and $-0.64$ according to \citet{2000AJ....120.1841F} and 
\citet{2003ApJ...592..906I}.  Establishing the origin of these deficiencies requires examining 
lighter $n$-capture elements.  In doing so, one must use care in dividing $n$-capture elements
into traditional $s$- and $r$-process categories.  Canonical ascriptions are usually based on 
solar-system contributions, and may not hold in other contexts.  Moreover, it has become
clear that a distinct light element primary process (LEPP) process is required to explain 
$n$-capture abundances in the Galactic field \citep{2004ApJ...601..864T}.    

\subsubsection{The heavy-to-light $n$-capture abundance ratio} 

As noted by \citet{2006MNRAS.370.2091C}, HD 134439/440 exhibit a peculiar {\it n}-capture 
abundance pattern: {\ }their mean [Ba/Y]${\sim}-0.10$ is smaller than the value of $+0.20$ exhibited 
by Galactic halo stars of similar metallicity (Table 5), and markedly lower than 
the mean value ($+0.60$) of the significant and real scatter evinced by dSph stars 
\citep{2003AJ....125..684S, 2004AJ....128.1177V, Tolstoy2009}.  Our stars are apparently deficient in heavy 
relative to light $n$-capture elements compared to the Galactic halo field and the majority of 
dSph systems studied to date.  

Figures 5 and 10 of \citet{2004ApJ...601..864T} indicate that $s$-contributions alone from AGB 
stars would result in [Ba/Y]$=+0.60$ to $+0.70$ at [Fe/H]$=-1.50$ in the Galaxy.  Indeed, several authors
\citep[e.g., ][]{2004AJ....128.1177V, Pompeia2008, Letarte2010} invoke such contributions to explain the scattered 
but clearly {\it super}solar [Ba/Y] values of dSph stars \citep[e.g., Figure 14 of][]{Tolstoy2009}.  
Such contributions seem to be 
missing (or muted by other contributions) in our two stars, which exhibit a {\it sub}solar ratio. This 
conclusion is strengthened by the upper limit on our stars' [Eu/${\alpha}$] ratio ($<+0.3$), which is 
lower than the super-Galactic dSph values \citep[Figure 19 of][]{Letarte2010} that have been ascribed to the action of AGB stars \citep{Letarte2010}. 

The lower [Ba/Y] ratio in our common proper motion pair is qualitatively consistent with the lower 
$r$-process component solar system Ba/Y ratio compared to the $s$-process component as seen in 
Figure 11 of \citet{Burris2000}, and suggests the importance of the action of the $r$-process in
our two stars' chemical history. While Ba is frequently 
referred to as an $s$-process element, its Galactic production at [Fe/H]$=-1.5$ is via the 
{\it r}-process \citep[Figure 5 of][]{2004ApJ...601..864T}, while the lighter $n$-capture element 
Y is produced via the {\it r}-process and the putative light element primary process 
\citep[LEPP;][]{2004ApJ...601..864T}.  The mean dSph ratio [Ba/Y]$\sim +0.60$ is in 
excellent agreement with the pure (i.e., no LEPP) {\it r}-process ratio predicted by Galactic 
chemical evolution models in Figure 12 of \citet{2004ApJ...601..864T}; indeed, these authors speculate 
that dSphs' super-Galactic pure {\it r}-process [Ba/Y] ratio might be driven by a deficiency in Y 
caused by loss of LEPP-rich ejecta to the ISM in dSph systems.  The lower [Ba/Y] ratio in 
134439/134440 might then be explained if the ISM of the parent dSph that birthed these 2 stars 
was somehow able to retain Y-rich LEPP products.  

However, we discount this possibility as follows.  The mean [Y/Fe] of our two stars only differs by 
$\sim$0.15 dex with respect to the mean Galactic halo field, a conclusion confirmed by additional and 
newer field data of \citet{Roed2010}, and lies in the midst of the scatter of 
dSph [Y/Fe] ratios \citep{2004AJ....128.1177V}. Our stars' [Ba/Fe] ratio, on the other hand, differs by 
0.4-0.5 dex with respect to the Galactic halo and is at the very bottom of the scattered dSph 
ratios \citep{2004AJ....128.1177V}.  It is thus Ba, and not Y, that is the source of our stars' 
unexpectedly low [Ba/Y] value. Inasmuch as the canonical association of Eu with the $r$-process 
seems secure and that the limited Galactic data suggest an $r$-process origin of Ag in metal-poor
stars \citep{1998AJ....116.2489C}, the $n$-capture abundances consistently suggest that 
our stars are deficient in heavy (relative to light) $r$-process products.  

Galactic halo stars evince a continuum of $r$-process abundance patterns rather than conforming
to a single standard pattern \citep[e.g.,][]{Aoki2000,Roed2010}.  Indeed, observed abundance distributions have 
identified a significant fraction of halo stars that are deficient in heavy $r$-process elements with respect to lighter ones.  
\citet{Boyd2012}  suggest that this ''truncated r-process'' may arise from nucleosynthesis via fall-back supernovae in which black
hole collapse terminates the $r$-process before it runs to completion and/or prevents the dispersal of heavy nuclides into
the interstellar medium.   However, the progenitor mass of such supernovae are believed to be $>25$ M$_{\odot}$ \citep{Heger92}.  This explanation for the low heavy-to-light $n$-capture ratio in HD 134439/440 is thus inconsistent 
with the evidence described above suggesting that these stars' nucleosynthetic history is dominated by Type II supernovae 
of low progenitor mass.  

The schematic Galactic $r$-process model of \citet{2001ApJ...559..925Q} divides Type II SN into two 
classes-- H and L.  The L class supernovae are less frequent events from lower mass stars that produce 
exclusively lighter (up to Ba) {\it n}-capture elements; H events are responsible for the production of 
heavier (A$\ge130$) {\it n}-capture elements, but produce light $n$-capture elements as well.  This 
phenomenological approach reproduces the Galactic $n$-capture abundance data well over the range 
$-3\le{\rm [Fe/H]}\le-1$ assuming no {\it s}-process contribution.  The relative L-event yields in 
Figure 2 of \citet{2001ApJ...559..925Q} produce the Ba/Ag and Ba/Y logarithmic number abundance 
ratios observed in our stars to within 0.17 dex; in contrast, their Figure 2 predicts Ba/Y ratios 
from H events some 0.80 dex larger than observed in our stars.  The agreement of the light 
{\it n}-capture abundance ratios in our stars with the L-event yields is consistent with our previous 
conclusion that abundances in our stars are dominated by nucleosynthetic products from lower mass 
Type II events; the observed deficiency in some heavy $n$-capture elements in our stars is also
consistent with this notion.  Additional interpretation of the HD 134439/440 $n$-capture abundances in 
the context of the \citet{2001ApJ...559..925Q} L,H-event model will require measurements of other 
possibly highly underabundant heavy $n$-capture elements, and computational 
exploration of possible L-event overflow that might allow the {\it r}-process to reach beyond 
A$\ge130$ and explain the detection of any such elements.   

A final possible explanation of the {\it n}-capture ratios in our stars that can also account 
for their unexpectedly high [Ca,Ti/Fe] ratios is the $\alpha$-process in a high neutron excess 
environment--a scenario originally envisioned by \citet{1960ApJ...132..565H} and first explored in detail by 
\citet{1992ApJ...395..202W}.  The latter authors show that, with sufficient neutron excess, the 
familiar $\alpha$-rich freeze-out ``merges'' into what may be an (or {\it the}) {\it r}-process.   This 
theoretical result is consistent with the observational work of \citep{Roed2010}, who detect the 
signatures of such production in the [Y/Eu] versus [Eu/Fe] relation evinced by metal-poor Galactic field 
stars, and conclude that the r-process nucleosynthesis is "ubiquitous" in all Type II SN.   
  
This process also naturally overproduces both Ca and Ti relative to other $\alpha$-elements--a signature
observed in HD 134439/440 when their abundances are compared to low mass Type II progenitor yields.  
Moreover, the production factors calculated under two different assumptions for model zone integration 
in Figure 2 of \citet{1992ApJ...395..202W} predict (heavy-to-light $n$-capture) [Ag/Y] ratios that bracket the observed ratio of our stars.  While nucleosynthesis in such a high-entropy environment is a promising explanation for the curious Ca, Ti, and {\it n}-capture abundances in our stars, a rigorous examination of 
this hypothesis requires abundance determinations of additional {\it n}-capture elements in our 
stars and $\alpha$-process calculations extended to elements with atomic mass beyond Ag. 

\section{SUMMARY}

CNO, Be, and Ag abundances are measured from high-resolution near-UV Keck/HIRES spectra for the 
metal-poor dwarf common proper motion pair HD 134439 and HD 134440.  The CNO abundance ratios are 
markedly sub-solar: [C/Fe]$=-0.56{\pm}0.09$, [N/Fe]$=-0.52{\pm}0.09$, and [O/Fe]$=-0.26{\pm}0.13$.  
The abundance ratios of 22 elements demonstrate a correlation with condensation temperature in our 
stars that is significant at the 95\% confidence level. Furthermore, the differences between these 
abundance ratios and those characterizing the Galactic halo field at similar [Fe/H] demonstrate 
correlation with $T_{\rm C}$ significant at the ${\ge}99$\% confidence level-- suggesting that the 
HD 134439/440 formation environment or these stars' post-formation chemical evolution differ remarkably from typical stellar denizens of the Galactic halo field. 

The correlation of abundance ratios with $T_{\rm C}$ in HD 134439/440 is opposite to the general 
trend seen in the gas-phase ISM, and instead reminiscent of a volatile-depleted refractory abundance 
pattern like that exhibited by primitive chondrites (e.g., Figure 1a of \citet{2005ASPC..341..632Y}).  
This pattern complements the hypothesis of \citet{2003ApJ...598L..47S}, who suggest low 
[${\alpha}/{\rm Fe}$] halo stars originate from photospheric accretion of planetessimal material.  
These authors invoke stellar accretion of circumstellar Fe-rich material that would drive down $\alpha$/Fe 
ratios to observed levels. While the HD 134439/440 photospheric abundance versus $T_{\rm C}$ relation 
is qualitatively consistent with the accretion of volatile-depleted refractory-rich material, we view this 
expalanation as unrealistically remarkable due to the required fine tuning of various parameters associated 
with the accretion process that would yield indistinguishable abundances in both components of our stellar pair. 

The $UVW$ motions and low values of [O,Mg,Si/Fe] for HD 134439/134440 provide kinematic and 
chemical evidence of an extra-Galactic origin for this common proper motion pair--the low 
${\alpha}$-element ratios being consistent in particular with an origin in a low/irregular star 
formation rate environment.  \citet{2006MNRAS.370.2091C} suggest that the abundance-$T_{\rm C}$ 
correlation exhibited by HD 134439/134440 can be explained by formation in an unusually dusty dSph 
environment.  We suggest, however, that the stars' Be abundances argue against this, and that the  
correlation is an illusory one having a less remarkable nucleosynthetic explanation.

The majority of HD 134439/134440 abundance ratios are consistent with production from Type II SNe having
low mass progenitors and/or recent chemical evolution models of dSph systems.  Indeed, the $Z=0$, 
${\sim}14{\ }{\rm M}_{\odot}$ Type II SNe progenitor yields of \citet{2006ApJ...653.1145K}) are able to match 
most of the stellar pair's observed abundance ratios; there is no evidence of Type Ia 
or low-to-intermediate mass star contributions to the abundance patterns.  These conclusions are consistent 
with those reached by \citet{2003AJ....125..684S} and \citet{2003AJ....125..707T} from {\it in situ} dSph abundance patterns.  

The low heavy-to-light $n$-capture abundance ratio of [Ba/Y]$\sim-0.10$ in HD 134439/440 motivated our study of 
$n$-capture elements. We estimate [Ag/Fe]$=-0.16{\pm}0.10$ and argue that [Eu/Fe] is significantly lower 
than in the similarly cool and metal-poor halo field dwarf HD 103095.  Taken together, the Ba, Ag, and Eu 
abundances consistently suggest the heavy-to-light $n$-capture element ratio is some ${\sim}$0.3-0.4 dex 
lower in HD 134439/134440 than in the mean Galactic halo field.  

We confront the $n$-capture data with 
expectations from three theoretical scenarios:{\ \ }a) the truncated $r$-process, {\ }b) the nucleosynthetic history 
of  HD 134439/440's precursor material being dominated by so-called L class events that produce mainly light $n$-capture elements in the phenomenological 2-component Type II SNe model of \citet{2001ApJ...559..925Q}, and 
{\ }c) a significant nucleosynthetic contribution to HD 134439/134440 from material produced in an 
$\alpha$-rich freeze-out process in a high neutron excess environment \citep{1992ApJ...395..202W}.  The 
truncated $r$-process requires supernovae of progenitor mass higher than is compatible with the non-$n$-capture 
abundances in HD 134439/440.   While scenarios b) and c) are consistent with the currently limited 
$n$-capture data, the latter is particularly 
attractive in that it also provides an excess production mechanism for Ca and Ti consistent with the 
observation that [Ca,Ti/Fe]$>$[O,Mg,Si/Fe] in our stars.   Further constraints will require abundance 
determinations of additional heavy $n$-capture elements in our stars, and additional theoretical predictions 
concerning the production of heavy $n$-capture elements in these two models that can be compared to future data. 

\acknowledgments
This work was supported by NSF grants AST 02-39518 and 09-08342 to J.R.K. and AST 05-05899 to A.M.B.

{\it Facility:} \facility{KECK}.

\clearpage

%input tables here
%\input{tab1.txt}
\begin{deluxetable}{l c c}
\tablewidth{0 pt}
\tablecaption{Observational Data}
\startdata
\hline
\hline
		& HD 134439	& HD 134440	\\
\hline
RA (2000)	& 15:10:13.09	& 15:10:12.97	\\
Dec (2000)	& -16:22:45.85	& -16:27:46.52	\\
V		& 9.07		& 9.44		\\
B-V		& 0.77		& 0.85		\\
Date (MJD)	& 53905.314698	& 53905.325741	\\
Exposure (s)	& 900		& 900		\\
S/N (CH)	& 192		& 173		\\
S/N (NH)	& 83		& 68		\\
S/N (OH)	& 87		& 70		\\
$\pi$\tablenotemark{a} (mas) & 34.14 & 33.68	\\
M\tablenotemark{b} ($M_{\odot}$) & 0.57  & 0.54 \\
M$_{V}$		& 7.076		& 7.392		\\
U\tablenotemark{b} (km/s)& $-310$ & $-311$      \\
V\tablenotemark{b} (km/s)& $-467$ & $-473$      \\
W\tablenotemark{b} (km/s)& $-44$  & $-48$       \\

\enddata

\tablenotetext{a}{From \citet{1997A&A...323L..49P}}
\tablenotetext{b}{From Table 6 of \citet{1994AJ....107.2240C}}
\label{tab:observation}
\end{deluxetable}
\clearpage

\begin{deluxetable}{l c c}
\tablewidth{0 pt}
\tablecaption{Stellar Parameters}
\startdata
\hline
\hline
Parameters				& HD 134439		& HD 134440		\\
\hline
[Fe/H]\tablenotemark{a} (dex)		& $-1.47{\pm}0.07$      & $-1.53{\pm}0.09$      \\
T$_{{\rm eff}}$\tablenotemark{a} (K)	& 5000 $\pm$ 50		& 4785 $\pm$ 50		\\
$\log g$ (dex)				& 4.69 $\pm$ 0.10	& 4.71 $\pm$ 0.10	\\
$\xi$\tablenotemark{a} (km/s)		& 1.5 $\pm$ 0.3		& 1.5 $\pm$ 0.3		\\
\enddata

\tablenotetext{a}{Values from \citet{1997AJ....113.2302K}.}

\label{tab:parameters}
\end{deluxetable}
\clearpage

\begin{deluxetable}{l r r r}
\tablewidth{0 pt}
\tablecaption{${\lambda}$4324 CH Region Linelist}
\startdata
\hline\hline
Wavelength & Species\tablenotemark{a} & ${\chi}_{\rm low}{\ }$ & {\ }log $gf$ \\
{\ \ \ \ \ \ }{\AA}  &    & eV{\ \ }       &      \\
\hline
4322.001    &  106.0	&  2.207    & -1.348 \\
4322.002    &	23.0	&  1.195    & -5.661 \\
4322.020    &  106.0	&  2.232    & -0.922 \\
4322.023    &	23.1	&  1.679    & -4.994 \\
4322.040    &	44.0	&  1.734    &  0.650 \\
4322.095    &  106.0	&  2.207    & -1.262 \\
4322.102    &  106.0	&  3.025    & -2.069 \\
4322.103    &	24.1	& 11.667    & -2.248 \\
4322.115    &	22.0	&  2.175    & -1.904 \\
4322.148    &	23.0	&  2.565    & -1.969 \\
4322.179    &  106.0	&  0.033    & -10.142 \\
4322.193    &	64.1	&  0.382    & -1.707 \\
4322.282    &	28.1	& 11.969    & -1.801 \\
4322.306    &  106.0	&  2.438    & -2.252 \\
4322.318    &	27.0	&  3.687    & -3.269 \\
\enddata
\tablenotetext{a}{In Tables 3-7, the species is identified by atomic 
number to the left of the decimal point, and ionization state (`0' is 
neutral, `1' is singly ionized, etc) to the right.  Diatomic molecules 
are denoted with both components' atomic numbers (`106' is CH, `607' 
is CN, etc).} 
\tablecomments{A machine-readable version of the entire table is available in the online journal.  A portion is shown here for guidance concerning form and content.}
\label{tab:chlines}
\end{deluxetable}
\clearpage

\begin{deluxetable}{l r r r}
\tablewidth{0 pt}
\tablecaption{${\lambda}3330$ NH Region Linelist}
\startdata
\hline\hline
Wavelength & Species & ${\chi}_{\rm low}{\ }$ & {\ }log $gf$ \\
{\ \ \ \ \ \ }{\AA}  &    & eV{\ \ }        &      \\
\hline
3325.001  &   44.0 &   0.385 &  -1.780 \\ 
3325.007  &   26.1 &   3.967 &  -2.707 \\
3325.010  &  107.0 &   0.445 &  -3.074 \\
3325.021  &   29.1 &  13.432 &  -3.543 \\
3325.037  &   26.1 &   8.923 &  -4.203 \\
3325.042  &  107.0 &   0.735 &  -0.949 \\
3325.050  &   58.1 &   0.446 &  -1.033 \\
3325.066  &  107.0 &   0.004 &  -5.459 \\
3325.074  &  107.0 &   0.735 &  -2.573 \\
3325.119  &  107.0 &   2.121 &  -0.598 \\
3325.121  &   90.1 &   0.514 &  -0.593 \\
3325.133  &   42.0 &   2.836 &  -0.949 \\
3325.148  &   22.0 &   2.134 &  -0.838 \\
3325.165  &  107.0 &   0.735 &  -2.037 \\
3325.180  &  107.0 &   2.121 &  -3.581 \\
\enddata
\tablecomments{A machine-readable version of the entire table is available in the online journal.  A portion is shown here for guidance concerning form and content.}
\label{tab:nhlines}
\end{deluxetable}
\clearpage 

\begin{deluxetable}{l r r r}
\tablewidth{0 pt}
\tablecaption{${\lambda}3130$ OH Region Linelist}
\startdata
\hline\hline
Wavelength & Species & ${\chi}_{\rm low}{\ }$ & {\ }$gf$ \\
{\ \ \ \ \ \ }{\AA}  &    & eV{\ \ }        &      \\
\hline
3128.060  &    108.0  &  0.541 & 3.758E-03 \\
3128.101  &    108.0  &  0.210 & 1.309E-03 \\
3128.154  &    108.0  &  1.598 & 1.014E-04 \\
3128.166  & 	25.1  &   6.91 & 1.901E-03 \\
3128.172  & 	25.0  &   3.86 & 2.280E-04 \\
3128.172  & 	25.1  &   8.77 & 2.415E-03 \\
3128.189  & 	23.0  &   1.85 & 1.660E-06 \\
3128.237  &    108.0  &  0.442 & 4.753E-04 \\
3128.269  & 	21.1  &   3.46 & 6.745E-01 \\
3128.286  &    108.0  &  0.210 & 1.035E-02 \\
3128.289  &    108.0  &  0.442 & 7.278E-04 \\
3128.302  & 	24.0  &   2.71 & 6.442E-07 \\
3128.304  & 	23.1  &   2.38 & 1.340E-01 \\
3128.356  & 	24.1  &   9.22 & 3.381E-04 \\
3128.356  &    108.0  &  1.712 & 2.455E-03 \\
\enddata
\tablecomments{A machine-readable version of the entire table is available in the online journal.  A portion is shown here for guidance concerning form and content.}
\label{tab:ohlines1}
\end{deluxetable}
\clearpage 

\begin{deluxetable}{l r r r}
\tablewidth{0 pt}
\tablecaption{${\lambda}3140$ OH Region Linelist}
\startdata
\hline\hline
Wavelength & Species & ${\chi}_{\rm low}{\ }$ & {\ }log $gf$ \\
{\ \ \ \ \ \ }{\AA}  &    & eV{\ \ }        &      \\
\hline
3138.014 &    26.0 &   2.469 &  -3.179	 \\
3138.036 &     2.0 &   0.616 &  -12.732  \\
3138.041 &   108.0 &   2.397 &  -1.787   \\
3138.055 &    23.1 &   3.759 &   0.333	 \\
3138.061 &    26.1 &   8.971 &   0.120	 \\
3138.088 &   108.0 &   1.098 &  -3.107   \\
3138.100 &   108.0 &   1.708 &  -1.574   \\
3138.110 &    26.1 &   9.053 &  -3.191	 \\
3138.118 &    28.1 &   8.522 &  -2.514	 \\
3138.138 &    25.0 &   2.920 &  -3.166	 \\
3138.165 &    23.0 &   0.287 &  -5.258	 \\
3138.178 &    26.1 &   8.256 &  -0.509	 \\
3138.183 &    25.1 &   6.111 &  -2.137	 \\
3138.192 &    24.0 &   3.122 &  -0.859	 \\
3138.198 &   108.0 &   2.330 &  -1.907   \\
\enddata
\tablecomments{A machine-readable version of the entire table is available in the online journal.  A portion is shown here for guidance concerning form and content.}
\label{tab:ohlines1}
\end{deluxetable}
\clearpage 

\begin{deluxetable}{l r r r}
\tablewidth{0 pt}
\tablecaption{${\lambda}3167$ OH Region Linelist}
\startdata
\hline\hline
Wavelength & Species & ${\chi}_{\rm low}{\ }$ & {\ }log $gf$ \\
{\ \ \ \ \ \ }{\AA}  &    & eV{\ \ }        &      \\
\hline
3164.042 &  25.1 &   4.82 & -5.23 \\ 
3164.057 &  24.0 &   3.37 & -2.68 \\ 
3164.071 &  26.1 &   8.99 & -4.22 \\ 
3164.078 &  26.0 &   3.37 & -3.53 \\ 
3164.092 &  25.1 &   5.99 & -4.16 \\ 
3164.123 &  27.0 &   2.14 & -3.87 \\ 
3164.134 &  26.0 &   3.42 & -4.12 \\ 
3164.141 &  23.1 &   2.76 & -3.22 \\ 
3164.153 &  58.1 &   0.30 & -0.25 \\ 
3164.159 &  28.0 &   1.95 & -2.56 \\ 
3164.164 &  24.1 &   9.23 & -3.79 \\ 
3164.172 &  25.1 &   7.76 & -4.05 \\ 
3164.256 &  24.1 &   4.07 & -2.11 \\ 
3164.267 &  26.1 &   3.89 & -3.19 \\ 
3164.296 &  26.0 &   2.45 & -4.50 \\ 
\enddata
\tablecomments{A machine-readable version of the entire table is available in the online journal.  A portion is shown here for guidance concerning form and content.}
\label{tab:ohlines1}
\end{deluxetable}
\clearpage 

\begin{deluxetable}{c c c c c}
\tablewidth{0 pt}
\tablecaption{Carbon Abundances}

\startdata
\hline\hline

           &\multicolumn{2}{c}{\underline{HD 134439}} &\multicolumn{2}{c}{\underline{HD 134440}} \\

$\lambda$  & $\log N$(C) &	[C/H]\tablenotemark{a}       & $\log N$(C) & [C/H]\tablenotemark{a} \\
{\AA}      & dex      &	  dex                        & dex      & dex    \\

\hline

4323.02    &  6.52    &  $-$1.87  		     &  6.20	&  $-$2.19 \\
4323.22    &  6.53    &  $-$1.86  		     &  6.21	&  $-$2.18 \\
4323.50    &  6.55    &  $-$1.84  		     &  6.24	&  $-$2.15 \\
4323.85    &  6.47    &  $-$1.92  		     &  6.09	&  $-$2.30 \\
4324.12    &  6.52    &  $-$1.87  		     &  6.22	&  $-$2.17 \\
4324.40    &  6.46    &  $-$1.93  		     &  6.11	&  $-$2.28 \\
4324.81    &  6.36    &  $-$2.03  		     &  6.09	&  $-$2.30 \\

\enddata

\tablenotetext{a}{Assumed solar carbon abundance of 
$\log N({\rm C})_{\odot}=8.39 \pm 0.05$ dex \citep{2005ASPC..336...25A}.}

\label{tab:carbon}
\end{deluxetable}
\clearpage

\begin{deluxetable}{c c c c c}
\tablewidth{0 pt}
\tablecaption{Nitrogen Abundances}

\startdata
\hline\hline

          &\multicolumn{2}{c}{\underline{HD 134439}} &\multicolumn{2}{c}{\underline{HD 134440}} \\

$\lambda$  & $\log N$(N) &	[N/H]\tablenotemark{a}       & $\log N$(N) & [N/H]\tablenotemark{a}  \\
{\AA}      & dex      &	dex                          & dex      & dex  \\

\hline

3326.41    &  5.84    &  $-$1.94 		      &  5.76    &  $-$2.02 \\
3328.20    &  5.84    &  $-$1.94		      &  5.65    &  $-$2.13 \\
3330.27    &  5.79    &  $-$1.99 		      &  5.61    &  $-$2.17 \\
3330.92    &  5.83    &  $-$1.95 		      &  5.70    &  $-$2.08 \\
%3325.71\tablenotemark{b}    &  5.93    &  $-$1.85 		      &  5.66    &  $-$2.12 \\
%3329.75\tablenotemark{b}    &  5.89    &  $-$1.89 		      &  5.69    &  $-$2.09 \\

\enddata

\tablenotetext{a}{Assumed solar nitrogen abundance of 
$\log N({\rm N})_{\odot}=7.78 \pm 0.06$ dex \citep{2005ASPC..336...25A}.}

\label{tab:nitrogen}
\end{deluxetable}
\clearpage

\begin{deluxetable}{c c c c c}
\tablewidth{0 pt}
\tablecaption{Oxygen Abundances}

\startdata
\hline\hline

           &\multicolumn{2}{c}{\underline{HD 134439}} &\multicolumn{2}{c}{\underline{HD 134440}} \\

$\lambda$  & $\log N$(O) &	[O/H]\tablenotemark{a}       & $\log N$(O) & [O/H]\tablenotemark{a}  \\
{\AA}      & dex      &	 dex                         &  dex     &  dex   \\

\hline

3129.94    &  7.11    &  $-$1.58  		     &  6.98	&  $-$1.71 \\
3138.78	   &  6.99    &  $-$1.70                       &  6.76	&  $-$1.93 \\
3138.92	   &  6.93    &  $-$1.76                       &  6.64	&  $-$2.05 \\
3139.30    &  7.07    &  $-$1.62  		     &  6.73	&  $-$1.96 \\
3140.51    &  7.06    &  $-$1.63  		     &  6.95	&  $-$1.74 \\
3141.66    &  6.96    &  $-$1.73  		     &  6.93	&  $-$1.76 \\
3167.17    &  7.03    &  $-$1.66                       &  6.79	&  $-$1.90 \\
\enddata

\tablenotetext{a}{Assumed solar oxygen abundance of 
$\log N({\rm O})_{\odot}=8.69 \pm 0.05$ dex \citep{2001ApJ...556L..63A}.}
\label{tab:oxygen}
\end{deluxetable}
\clearpage

\begin{deluxetable}{c c c c c c c}
\tablewidth{0 pt}
\tablecaption{Abundance Sensitivities and Uncertainties}
%\rotate
%\tabletypesize{\scriptsize}

\startdata
\hline\hline
   
%         & \multicolumn{7}{l}{\underline{HD 134439}} & \multicolumn{7}{l}{\underline{HD 134440}} &   \\
         & $\Delta$T$_{{\rm eff}}$ & $\Delta\log g$  &  $\Delta\xi$  & $\Delta cont$\tablenotemark{a} & $\sigma$\tablenotemark{b} & $\sigma_{avg}$\tablenotemark{c} \\
%         & $\Delta$T$_{eff}$ & $\Delta$log $g$  &  $\Delta\xi$\tablenotemark{b}  & $\Delta$[Fe/H] & $\Delta$cont & $\Delta\chi^{2}$ fit & $\sigma_{tot}$ 
         & $\pm$150 K & $\pm0.20$ dex & $\pm0.5$ km/s & dex & dex & dex  \\

\hline

$\Delta$[C/Fe] & $\pm$0.09 & $\mp$0.04 & $\pm$0.03 & 0.04 & 0.07 & 0.09  \\

$\Delta$[N/Fe] & $\pm$0.12 & $\mp$0.05 & $\pm$0.02 & 0.06 & 0.05 & 0.09  \\

$\Delta$[O/Fe] & $\pm$0.15 & $\mp$0.06 & $\pm$0.00 & 0.06 & 0.10 & 0.13  \\

\enddata

\tablenotetext{a}{This is the abundance uncertainty arising from the uncertainty in setting the pseudo-continuum of the spectra.}
\tablenotetext{b}{$\sigma$ is the statistical uncertainty in the mean abundance from the molecular lines measured.}
\tablenotetext{c}{This is the final total uncertainty in the [X/Fe] ratios for HD 134439 and HD 134440.} 
\tablecomments{We have adopted the sensitivities of [Fe/H] due to T$_{{\rm eff}}$, 
$\log g$, $\xi$, and statistical uncertainty from Table 3 of \citet{1997AJ....113.2302K} 
for the calculations in this table.}

\label{tab:sensitivities}
\end{deluxetable}
\clearpage

\begin{deluxetable}{c c c c c c c}
\tablewidth{0 pt}
\tablecaption{Elemental Abundances}

\startdata
\hline\hline
	 & X    &[X/Fe]		       &[X/Fe]  	       &[X/Fe]$_{avg}$\tablenotemark{a}  & [X/Fe]$_{\rm Halo}$  & T$_{{\rm c}}$(K)\tablenotemark{b} \\
	 &      & HD 134439  		       & HD 134440      & & \@[Fe/H]$=-1.5$  &    \\
\hline
         &C     &$-$0.43 		       &$-$0.70		       &$-$0.56 $\pm$0.09  &$-0.12$ &40   \\  
         &N	&$-$0.48  	       &$-$0.57		       &$-$0.52 $\pm$ 0.09         &$+0.00$ &123  \\
				    			     						 \\
	 &O	&$-$0.20		       &$-$0.33		       &$-$0.26 $\pm$ 0.13 &$+0.54$ &180  \\  
$\alpha$ &Mg	&$-$0.09     	       &$-$0.10   	       &$-$0.10 $\pm$ 0.05	   &$+0.39$ &1336 \\
elements &Si	&$+$0.04     	       &$+$0.04   	       &$+$0.04 $\pm$ 0.10	   &$+0.30$ &1310 \\ 
	 &Ca	&$+$0.09     	       &$+$0.08   	       &$+$0.08 $\pm$ 0.04	   &$+0.31$ &1517 \\  
	 &Ti	&$+$0.18     	       &$+$0.20   	       &$+$0.19 $\pm$ 0.09	   &$+0.24$ &1582 \\  
		           		        							 \\
	 &Na	&$-$0.48     	       &$-$0.49   	       &$-$0.48 $\pm$ 0.06	   &$-0.16$ &958  \\  
	 &Al	&$-$0.22     	       &$-$0.19   	       &$-$0.20 $\pm$ 0.04	   &$-0.31$ &1653 \\ 
	 &K	&$+$0.10     	       &$+$0.08   	       &$+$0.09 $\pm$ 0.08	   &$+0.50$ &1006 \\  
odd Z 	 &Sc	&$-$0.03     	       &$-$0.04   	       &$-$0.04 $\pm$ 0.07	   &$-0.09$ &1659 \\  
elements &V	&$+$0.01     	       &$+$0.05   	       &$+$0.03 $\pm$ 0.08	   &$+0.08$ &1429 \\  
	 &Mn	&$-$0.41     	       &$-$0.37   	       &$-$0.39 $\pm$ 0.05	   &$-0.45$ &1158 \\  
	 &Co	&$-$0.03     	       &$+$0.05   	       &$+$0.01 $\pm$ 0.05	   &$-0.11$ &1352 \\  
	 &Cu	&$-$0.69     	       &\nodata		       &\nodata			   &$-0.57$ &1037 \\
													 \\
	 &Fe\tablenotemark{c}&($-$1.47)  &($-$1.53)		       &($-$1.50 $\pm$ 0.09) & $+0.$ &1334 \\
even Z	 &Cr	&$-$0.01     	       &$+$0.06   	       &$+$0.02 $\pm$ 0.06	   &$-0.04$  &1296 \\  
elements &Ni	&$-$0.12     	       &$-$0.13   	       &$-$0.12 $\pm$ 0.05	   &$-0.06$  &1353 \\  
	 &Zn	&$-$0.04     	       &$-$0.06   	       &$-$0.05 $\pm$ 0.10         &$+0.06$  &726  \\  
		           		        							 \\
n$-$capture&Y	&$-$0.22     	       &$-$0.28   	       &$-$0.25 $\pm$ 0.08	   &$-0.07$  &1659 \\  
elements &Ag	&$-$0.19		       &$-$0.13		       &$-$0.16 $\pm$ 0.10 &$+0.28$  &996	 \\
	 &Ba	&$-$0.35     	       &$-$0.36   	       &$-$0.36 $\pm$ 0.07	   &$+0.05$  &1455 \\
\enddata

\tablenotetext{a}{[X/Fe]$_{avg}$ is the average [X/Fe] of HD 134439 and HD 134440.}
\tablenotetext{b}{Condensation temperatures are taken from Table 8 of \citet{2003ApJ...591.1220L}.}
\tablenotetext{c}{These are [Fe/H] values from \citet{1997AJ....113.2302K}.}
\tablecomments{Elemental abundances and uncertainties, with the exception of C, N, O, Fe, and Ag are 
taken from \citet{2006MNRAS.370.2091C}.}

\label{tab:abundances}
\end{deluxetable}
\clearpage

%input figures here

\end{document}